\documentclass[%
 reprint,
 amsmath,amssymb,
 aps,
prl,
]{revtex4-2}

\usepackage{graphicx}
\usepackage{dcolumn}
\usepackage{bm}
\usepackage[mathlines]{lineno}

\begin{document}
	\eprint{arXiv:2111.05438 (2021)}


\title{Parametric mechanism of the magnetization reversal  as a low- power recording mechanism for MRAM. Measurement of spin- accumulation- induced in-plane magnetic field in a FeB nanomagnet. }

\author{Vadym Zayets}
\affiliation{National Institute of Advanced Industrial Science and Technology (AIST), Umezono 1-1-1, Tsukuba, Ibaraki, Japan}


\begin{abstract}
The parametric torque presents a promising recording mechanism for MRAM, complementing Spin Transfer Torque and Spin Orbit Torque, enabling magnetization reversal in a nanomagnet using a DC electrical current. Its resonance nature allows for optimization of magnetization reversal at a lower current, presenting an opportunity for a lower recording current and, therefore, for efficient and high-performance operation in modern MRAM technology. The in-plane magnetic field generated by spin accumulation serves as the driving force behind this torque. Experimental measurements of the current-induced in-plane magnetic field in the FeB nanomagnet reveal its magnitude to be around 40 Gauss  at a current density of 25 mA/$\mu m^2$, a value adequate for facilitating parametric magnetization reversal. The parametric torque is analytically calculated by solving the Landau- Lifshitz equation. Analytical calculations demonstrate its potential in advancing modern MRAM technology.
\end{abstract}


\maketitle


The primary challenge in the current development of MRAM lies in reducing the energy required for recording. Overcoming this challenge within conventional recording methods like Spin Torque (STT) and Spin-Orbit Torque (SOT) proves difficult due to fundamental limitations. These limitations arise from the minimum amount of injected spin-polarized electrons necessary to generate an adequate STT or SOT torque for magnetization reversal in a ferromagnetic layer \cite{MRAMCubukcu,FertpGMR,Flederling1999SpinInjectionLD,Crooker2005SpinInjectionImage,Ohno1999_SpinInjection,Crooker2005bImagingSpinTransport}.  As a result, the required recording current remains relatively high, impeding energy reduction efforts.

One potential solution to address the energy reduction in MRAM recording involves employing a resonance recording method. In this approach, a significantly smaller quantity of spin-polarized electrons produces an oscillating torque that resonates with Larmor oscillation. After each oscillation period, the torque incrementally increases the precession angle. Although modest, this resonance torque proves sufficient for achieving magnetization reversal. Consequently, employing this technique allows for a substantial reduction in the required recording current.

Achieving efficient parametric enhancement of an oscillation necessitates precise frequency and phase matching between the pump and oscillator. To accomplish this matching, external control of the pumping parameters and precise feedback tuning of phase and frequency must be implemented.

In the case of magnetization precession, the conditions for parametric enhancement of precession angle and magnetization reversal become even more stringent. This complexity arises from the dependence of precession phase and frequency on the precession angle \cite{ZayetsArch2021Parametric}, making it challenging to maintain resonance. Even if the pump is initially tuned to the precession frequency and a resonant increase in precession angle occurs, any further enhancement is hampered by the detuning caused by the angle increase, thereby suppressing additional enhancement.

For instance, in a conventional ferromagnetic-resonance (FMR) measurement\cite{FMR_Kittel}, a ferromagnetic sample is exposed to a microwave at the Larmor frequency. However, this illumination alone does not lead to magnetization reversal. Despite the initial resonant enhancement of precession, a frequency and phase mismatch between the microwave and magnetization precession arises after a slight increase in the precession angle. This mismatch disrupts the resonance pumping, preventing further augmentation of the precession angle.

To achieve further increase in precession angle through parametric pumping, which eventually would lead to the parametric magnetization reversal,   it is necessary to dynamically adjust the frequency and phase of the pumping to accommodate the shifting resonance conditions arising from the increasing precession angle. One potential solution is to incorporate real-time feedback using a parameter that is both proportional to the precession parameters and influences the efficiency of the pumping process. This approach allows for the adjustment of pumping parameters in response to the evolving resonance conditions, enhancing the overall effectiveness of the parametric pumping and promoting the desired magnetization reversal.

The necessary feedback loop can be established by utilizing magneto-resistance. A practical example involves the use of a magnetic tunnel junction (MTJ), where the resistance is proportional to the mutual angle between the magnetization directions of its two ferromagnetic layers\cite{MRAMCubukcu}.  Magnetization of one of these layers is pinned due to a strong magnetic anisotropy and remains fixed without any tilting. The magnetization of the second layer can be tilted, for example, by means of applying an external magnetic field.  The magneto-resistance of the MTJ, which is proportional to the tilt and subsequently the precession angle of the second layer, offers the crucial feedback required for effective parametric pumping of the magnetization precession.

When a DC voltage is applied to the MTJ,  the current flowing through it undergoes modulation in synchronization with the frequency and phase of magnetization precession. The amplitude of this current is directly linked to the precession angle. The establishment of a crucial feedback loop occurs when the electrical current induces the tilting of the magnetization, resulting in the precession pumping.  

The focus of the paper is the examination of the parametric mechanism of magnetization reversal, specifically attributed to the magnetic field induced by spin polarization. When an electrical current passes through a ferromagnetic nanomagnet, spin-polarized conduction electrons accumulate at the boundaries. This accumulated spin creates a magnetic field that tilts the magnetization. Given that the quantity of spin accumulation is directly proportional to the current, the magnetization tilt exhibits a proportional relationship to the current. This provides the necessary feedback between the current and precession pumping.

In Section 2, the experimental measurements of the spin polarization-induced magnetic field in a FeB nanomagnet are described. Section 3 describes  the mechanism by which the tilt of magnetization, induced by the spin- polarization-induced magnetic field, leads to parametric pumping of the magnetization precession and ultimately results in magnetization reversal. Section 4 provides detailed insights into the establishment of the feedback loop for parametric magnetization reversal in a MTJ, as well as the conditions under which parametric enhancement of magnetization precession occurs in this particular structure.

\section{Measurement of the magnetic field induced by spin- accumulated electrons in FeB nanomagnet}

Instead of using an MTJ where the magnetic field of the second magnetic layer could introduce a systematic error, the magnetic field resulting from the current-induced spin accumulation is measured in a single- layer FeB nanomagnet. The measurement process involved scanning an in-plane external magnetic field $H_x$ in two opposite directions, while the magnetization angle was determined using a Hall setup \cite{Zayets2022IEEE_Hci}. The in-plane magnetic field $H_x$ is directed perpendicular to the magnetic easy axis of the nanomagnet, thus causing the magnetization to tilt (See \ref{AppendixExpSetup}). The presence of an additional intrinsic in-plane magnetic field $H_{||}$ caused a constant shift along the x-axis in the linear relationship between the in-plane magnetization component $M_x$ and the applied in-plane magnetic field $H_x$: 

\begin{equation}\label{Hani}
	\frac{M_x}{M}=\frac{H_x+H_{||}}{H_{ani}}
\end{equation}

where $H_{ani}$ is the anisotropy field.

By performing a fit of the measured $M_x$  against $H_x$, the value of $H_{||}$  was determined as an offset in Eq. \ref{Hani} with a precision of approximately 0.5 Gauss. The two components of $H_{||}$, one parallel to the current direction and the other perpendicular to it, are measured by scanning the in-plane external magnetic field $H_x$ either along or perpendicular to the current direction. Details of the measurement technique are described in ref. \cite{Zayets2022IEEE_Hci} and \ref{AppendixExpSetup}. In addition, an external magnetic field $H_z$ was applied perpendicular to the plane and served as a parameter for the measurement. The nanomagnet under investigation was composed of a 1.2 nm-thick FeB layer grown on a 3-nm-thick tungsten  Hall bar. Equilibrium magnetization direction of the nanomagnet is perpendicular-to-plane.The dimensions of the nanomagnet are 3000 x 3000 nm.  In order to avoid potential measurement-related systematic errors, the experimental setup was thoroughly calibrated using the Hall angle measurement  in a non-magnetic metal.

\begin{figure}[t]
	\begin{center}
		\includegraphics[width=7 cm]{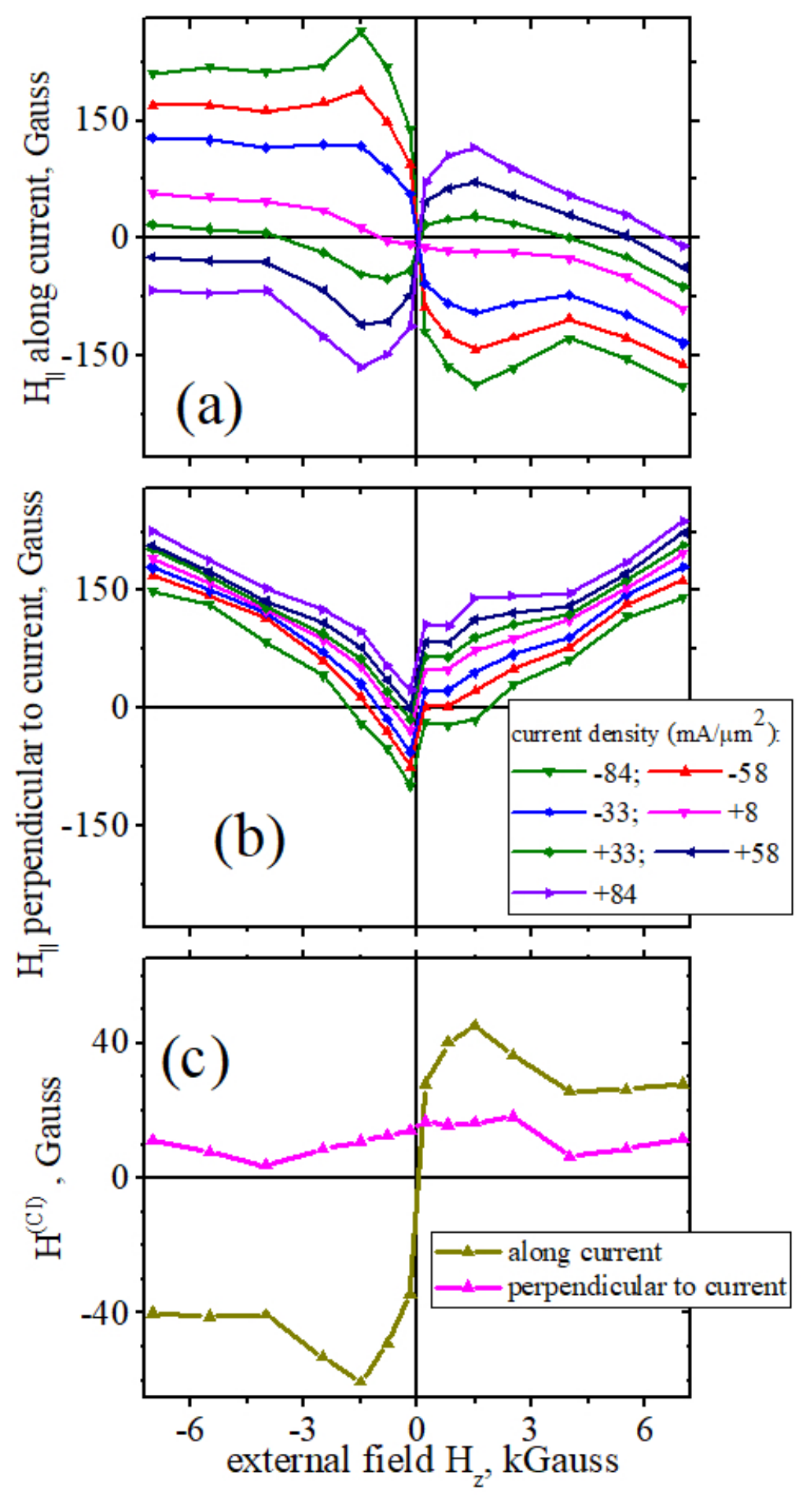}
	\end{center}
	\caption{\label{fig:fig1} 
		In-plane magnetic field $H_{||}$ measured in FeB nanomagnet (a) along current and (b) perpendicular- to- current as a function of external perpendicular- to plane magnetic field  $H_z$ at a different current density. (c) Current- induced in- plane magnetic field $H^{(CI)}$ at current density of 25 mA/$\mu m^2$, which is calculated as a difference between blue and pink lines of Figs. (a) (b)  
	}
\end{figure}


Figures \ref{fig:fig1}(a,b) depict the in-plane components of the magnetic field, $H_{||}$ measured in two directions: along and perpendicular to the current direction, at varying current densities. $H_{||}$ spans a range of -150 Gauss to +200 Gauss and is significantly influenced by the external magnetic field. 

Interestingly, even in the absence of an electrical current or with a weak current (as indicated by the red line), $H_{||}$ is still present. This implies that, in a state of equilibrium, the magnetization of the nanomagnet exhibits a slight tilt rather than being perfectly perpendicular to the plane. Remarkably, we observed a consistent tilt angle across almost a hundred FeB and FeCoB nanomagnets, each possessing different structures, compositions, and sizes ranging from 50 x 50 nm to 3000 x 3000 nm. These findings remained consistent throughout all the measurements conducted thus far.

The measured $H_{||}$ comprises all possible in-plane magnetic fields originating from various possible sources, which are applied to the nanomagnet resulting in the tilting of its magnetization. What makes $H_{||}$ particularly noteworthy is the existence of a substantial component directly proportional to the applied current. Figure \ref{fig:fig1}(c) shows  this current-induced magnetic field  $H^{(CI)}$ at  the current density of 33 mA/$\mu m^2$ versus external field $H_z$.This magnetic field plays a pivotal role in the parametric magnetization reversal process as it tilts the magnetization of the nanomagnet when a current passes through it. The current- induced magnetic field $H^{(CI)}$ is calculated as a difference of $H_{||}$ measured at 33 and 8 mA/$\mu m^2$.

It's worth noting that $H^{(CI)}$ is the same magnetic field responsible for generating the field-like torque (FLT) and its corresponding magnetization reversal mechanism\cite{SecondHarmonic_Garello2013,SecondHarmonic_Kim2013,FielLikeTorqueViennaTheory2015,FielLikeTorqueUKTheory2017}. In this mechanism \cite{FielLikeTorqueTheory2015}, $H^{(CI)}$ initiates magnetization precession. As the magnetization angle theta surpasses 90 degrees, $H^{(CI)}$ is switched off. As at this moment the magnetization angle is sufficiently large, the magnetization stabilizes in the reversed direction ($\theta =180^0$).

For the FLT reversal to take place, the critical requirement is a sufficiently large $H^{(CI)}$ to induce a magnetization tilt of at least $45^0$. Only under this condition can theta reach 90 degrees, enabling magnetization reversal\cite{FielLikeTorqueTheory2015}.  Across all our measured nanomagnets, $H^{(CI)}$ is consistently at least an order of magnitude smaller than the threshold required for the magnetization reversal by the field-like torque, even at a current density of 200 mA/$\mu m^2$, which is five times greater than that employed in MRAM. In contrast, the measured $H^{(CI)}$ is more than sufficient to facilitate the parametric magnetization reversal.

The in-plane magnetic field $H^{(CI)}$ can also be measured using the second harmonic (SH) method \cite{SecondHarmonic_Kim2013,SecondHarmonic_Pi2010,SecondHarmonic_Garello2013}. However, this approach is substantially influenced by notable systematic errors\cite{Zayets2022IEEE_Hci}, making it challenging to precisely determine the absolute value of this magnetic field. In addition to $H^{(CI)}$, the measured SH signal is impacted by various factors, such as current modulation of the anisotropy field\cite{MMM2020_Hani_Hoff,TMRC2021_2param,MMM2021_2param}, magnetization precession due to spin injection\cite{Zayets2022IEEE_Hci}, and current modulation of spin polarization in conduction electrons\cite{MMM2020_sp,Zayets2020MishenkoSpinPol}. These contributions collectively contribute to the overall systematic error of the SH method for the $H^{(CI)}$ measurement.The presence of this systematic error significantly complicates and effectively hinders the accurate determination of the absolute value of $H^{(CI)}$ in the SH method.

This magnetic field $H^{(CI)}$  is truly exceptional, as it tilts the magnetization of the nanomagnet when a current passes through it. There are several potential sources of this magnetic field. Each source can be distinguished by its distinct dependence on the magnetic field $H_z$. The conventional Oersted magnetic field is one candidate as it is created by an electrical current, circulates around a wire and can also tilt the magnetization.  Given that a portion of the current passes beneath the nanomagnet via the tungsten wire, it's plausible that the Oersted field contributes to the magnetization tilt in the nanomagnet. Notably, the Oersted magnetic field remains unaffected by the external magnetic field, enabling its identification and measurement as a consistent offset in Fig. 3. As is expected, the Oersted magnetic field only exists in the direction perpendicular to the current. It is estimated to be approximately 10 Gauss.

The spin accumulation emerges as another plausible source for the magnetic field $H^{(CI)}$, as it accounts for all observed characteristics of $H^{(CI)}$. The amount of the spin accumulation is proportional to the electrical current \cite{Crooker2005bImagingSpinTransport,SpinHallDyakonov1971,Flederling1999SpinInjectionLD,SpinHallKato2004}. Given that the accumulated spin-polarized conduction electrons possess a magnetic moment, their presence creates a magnetic field proportional to the current. Furthermore, the spin accumulation takes place at a nanomagnet's interface  and, therefore, its magnetic field does affect the spins in the bulk of the nanomagnet, causing the observed tilting of the nanomagnet's magnetization.

The spin precession and the alignment of spin along an external magnetic field are key features that aid in identifying the contribution of spin accumulation.  As there is a spin component perpendicular to this easy axis and the magnetic field $H_z$, spin precession and spin alignment occur along $H_z$ for the spin- accumulated electrons. This alignment effect leads to a reduction in the measured in-plane component of $H^{(CI)}$. As $H_z$ increases, the frequency of spin precession rises, and the alignment out of the in-plane direction accelerates.  Consequently, the $H^{(CI)}$ component associated with spin accumulation should exhibit oscillation and decay with the increase of $H_z$. This distinctive feature of spin accumulation is clearly evident in the measurement shown in Figure  \ref{fig:fig1}(c), where both components of $H^{(CI)}$, along and perpendicular to the current, display oscillation and decay patterns with $H_z$.

Another notable feature of these oscillations provides crucial insight into the origin of spin accumulation. When the magnetic field $H_z$ is reversed, the polarity of the oscillation remains unchanged for the component perpendicular to the current, but it reverses for the component along the current. This behavior implies that spin accumulation is  initially created with the spin direction perpendicular to the current.  Subsequently, due to spin precession, the spin rotates towards the direction along the current. When $H_z$ is reversed, the rotation direction also reverses, leading to the reversal of the along-current component of the oscillation while leaving the perpendicular-to-current component unaffected.

The creation of the spin- accumulation, when spin direction is parallel to surface and perpendicular to  the current direction, is a distinguished feature of the Spin Hall effect \cite{SpinHallKato2004,SpinHallSinolva2005}. Therefore, it is highly likely that the origin of the observed spin accumulation is indeed the Spin Hall effect \cite{SpinHallDyakonov1971,SpinHallHirsh1999}.

The third contribution to $H^{(CI)}$, which is directed along the current, remains independent of $H_z$ but undergoes a polarity reversal during magnetization reversal.The contribution corresponds to the offset of oscillation of the green line in Fig.\ref{fig:fig1}(c) and is evaluated to be about 40 Gauss. It neither decays nor oscillates with changing $H_z$, indicating its insensitivity to spin accumulation or $H_z$.  Its correlation with magnetization, evidenced by its reversal during magnetization reversal and independence of $H_z$, suggests its origin is related to the  magnetization. The origin of this contribution remains unidentified. Intriguingly, a comparable effect was observed in an antiferromagnetic system, where an electrical current affected spins of localized electrons without generating spin accumulation \cite{antiferroFMRonCurrent2021}.

Regardless of the specific origin of the measured magnetic field $H^{(CI)}$, it possesses two crucial characteristics that play a pivotal role in parametric magnetization reversal.  Firstly, the field exhibits a linear proportionality to the electrical current. Secondly, it causes a tilting of the nanomagnet magnetization.

\section{Mechanism of parametric magnetization reversal}

This section elucidates the mechanism behind how the modest magnetic field, induced by spin accumulation and directly proportional to the electrical current, can achieve magnetization reversal when the current modulation is accurately synchronized with the magnetization precession.

It is worth noting that the measured current-induced magnetic field $H^{(CI)}$ is relatively small, and an external magnetic field of the same strength is far from sufficient to reverse the nanomagnet's magnetization. In typical MRAM recording, the current density used is in range 10-50 mA/$\mu m$\cite{MRAMCubukcu}. At a current density of 25 mA/$\mu m$, the measured value of $H^{(CI)}$ is only about 40 $Gauss$ (as shown in Fig. \ref{fig:fig1} (c)).

The internal magnetic field, which maintains the magnetization perpendicular to the plane \cite{Johnson1996Hani,Zayets2020MishenkoSpinPol}, is approximately equivalent to the anisotropy field, measured to be around 5 $kGauss$ for this particular nanomagnet. When an external magnetic field of 40 $Gauss$ is applied perpendicular to the 5 $kGauss$ anisotropy field, it tilts the magnetization only very slightly, at an angle of approximately 450 $mdeg$. This tilting angle is too small to result in magnetization reversal \cite{ZayetsArch2019Hc}.

Magnetization reversal can take place under the influence of such a small magnetic field only when the conditions of parametric resonance are fulfilled. In this process, the weak magnetic field is modulated in resonance with the magnetization precession. As a consequence, the precession undergoes resonant enhancement, causing the precession angle to increase after each oscillation period until magnetization reversal eventually occurs.

\begin{figure}[t]
	\begin{center}
		\includegraphics[width=8 cm]{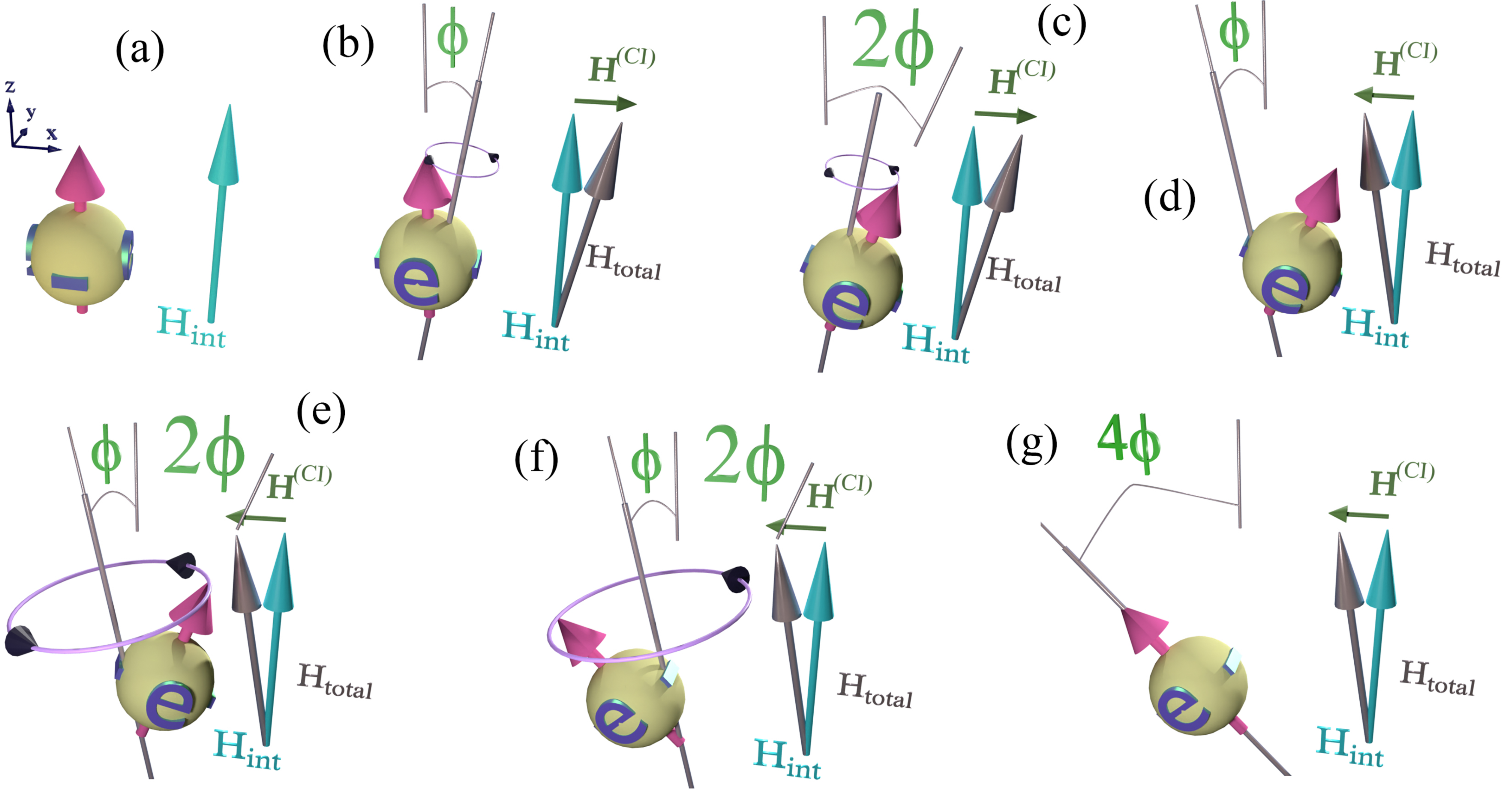}
	\end{center}
	\caption{\label{fig:fig2} 
		Parametric  enhancement of magnetization oscillation under oscillating magnetic field $H^{(CI)}$ (green arrow). (a) Conditions at equilibrium: magnetization (green ball) is directed along internal magnetic field $H_{int}$ (blue arrow). (b) When $H^{(CI)}$ is applied, the total magnetic field $H_{total}$ (grey arrow) turns away from the equilibrium direction at angle $\phi$ and magnetization precession starts; (c-f) precession angle increases due to synchronized change of $H^{(CI)}$ ; (g) precession angle becomes $4\phi$ after one period of precession. 
	}
\end{figure}

Figure \ref{fig:fig2} presents a schematic diagram elucidating the mechanism of magnetization precession enhancement through parametric resonance. In Fig. \ref{fig:fig2}(a), the initial states depict the nanomagnet's magnetization being perpendicular to the plane (along the z-axis, see \ref{AppendixExpSetup}) due to Perpendicular Magnetic Anisotropy (PMA)\cite{Johnson1996Hani}. An intrinsic magnetic field, $H_{int}$, aligned along the z-axis, maintains the magnetization along that axis. When an electrical current flows through the nanomagnet, it induces a magnetic field, $H^{(CI)}$, perpendicular to $H_{int}$ (along the x-axis in Fig. \ref{fig:fig2}(b)). The total magnetic field, $H_{total}$, applied to the nanomagnet results from the vector sum of $H_{int}$ and $H^{(CI)}$.

Upon application of $H^{(CI)}$, and the deviation of $H_{total}$ from the magnetization direction, the magnetization precession initiates around $H_{total}$. At this moment, the precession angle $\phi$ represents the angle between the magnetization and $H_{total}$. As was mentioned above, the angle  $\phi$ is very small, approximately a hundred millidegrees.

After half of the precession period, the magnetization rotates by an angle of $2\phi$ with respect to the z-axis (Fig. \ref{fig:fig2}(c)). At this point, the direction of $H^{(CI)}$ is reversed, leading to a reversal of the inclination of $H_{total}$ with respect to the z-axis. However, the absolute value of the inclination angle remains equal to $\phi$ (Fig. \ref{fig:fig2}(d)). Consequently, the angle between the magnetization and $H_{total}$ becomes $2\phi+\phi =3\phi $ (Fig. \ref{fig:fig2}(e)), resulting in the precession angle reaching $3\phi$.

During the subsequent precession, the angle between the magnetization and the z-axis continues to increase, reaching $3\phi+\phi =4\phi $ after half a period of the precession (Fig. \ref{fig:fig2}(h)). Thus, when the oscillation of $H^{(CI)}$ is synchronized with the magnetization precession, the precession angle increases by an angle of $4\phi$ at each precession period. Despite the initial small inclination angle $\phi$ the precession angle rapidly becomes large due to the parametric pumping of the precession.

In Appendix, the Landau-Leifshitz equation with an oscillating pumping magnetic field was analytically solved. Similar to the schematic depiction in Fig.  \ref{fig:fig2}, the analytical solution predicts a significant parametrically induced torque, even with a relatively small oscillating pumping magnetic field. However, this parametric torque diminishes rapidly when the frequency or phase of the oscillating pumping magnetic field becomes detuned from the magnetization precession.


\section{ Positive feedback loop for parametric reversal in MTJ}

As previously discussed, a small oscillating magnetic field can effectively enhance precession and efficiently reverse magnetization. However, precise synchronization between the pumping magnetic field and the magnetization precession is essential. This task is not straightforward since the frequency of the magnetization precession changes as the precession angle increases\cite{ZayetsArch2021Parametric}, necessitating dynamic tuning of the pump frequency.

As previously mentioned, a practical approach for real-time feedback is to incorporate the magneto-resistance, which varies with the magnetization precession and, therefore, is proportional to the precession  angle and frequency. To adapt to the changing resonance conditions, the pumping efficiency should be linked to the modulated magneto-resistance. This parametric pumping mechanism can be implemented in MTJ by utilizing the magnetic field $H^{(CI)}$ generated by the spin accumulation.

\begin{figure}[t]
	\begin{center}
		\includegraphics[width=8.5cm]{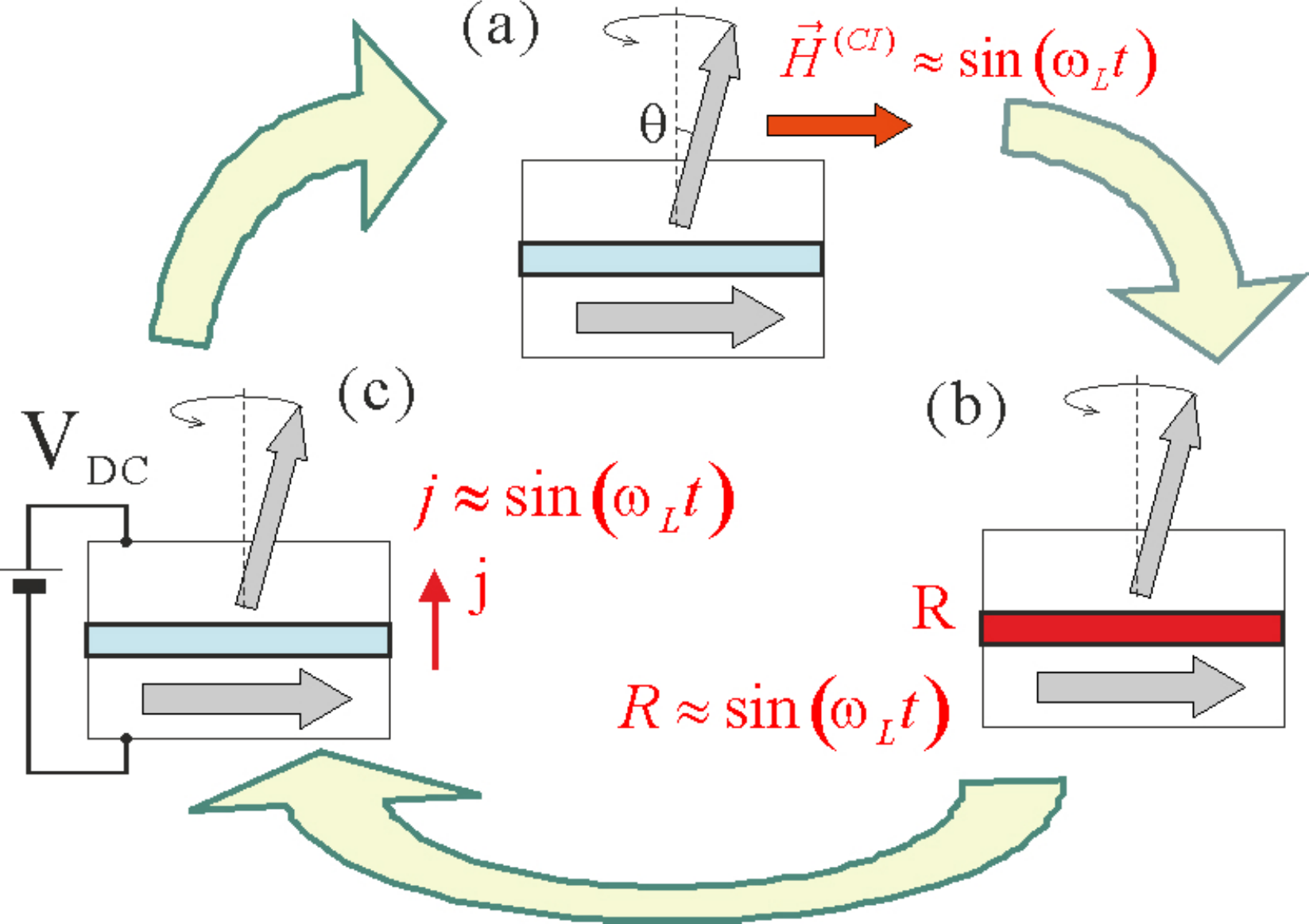}
	\end{center}
	\caption{\label{fig:fig3} 
		The positive feedback loop, which is critical  for the existence of the parametric torque. Gray arrows depict the magnetization of the "free" and "pin" layers of the MTJ. (a) An electrical current induced spin accumulation, leading to the magnetic field $H^{(CI)}$, which tilts the "free" layer's magnetization, increasing the precession angle $\theta$. (b) Precession modulates the MTJ resistance with frequency $\omega_L$, and a larger precession angle $\theta$ results in a larger modulation. (c) Under a DC voltage, the electrical current is modulated at $\omega_L$, where the larger resistance modulation leads to a larger current modulation. (a) The current modulates $H^{(CI)}$, resulting in a larger magnetization tilt and, consequently, a larger precession angle.
	}
\end{figure}


Figure \ref{fig:fig3} illustrates the positive feedback loop within the MTJ structure. In this setup, the equilibrium magnetization direction of the "pin" layer is in-plane, while the "free" layer's magnetization is perpendicular to the plane. The "pin" layer's magnetization is pinned due to strong magnetic anisotropy and remains fixed without any tilting. On the other hand, the magnetization precession of the "free" layer modulates the mutual angle between the magnetizations of both layers, consequently modulating the resistance of the MTJ (Fig. \ref{fig:fig3}(b)).

Under the application of a DC voltage to the MTJ, the electrical current, which is inversely proportional to the MTJ resistance, undergoes modulation at the precession frequency $\omega_L$ (Fig. \ref{fig:fig3}(c)). The electrical current creates the spin accumulation, which creates magnetic field $H^{(CI)}$. Since $H^{(CI)}$ is linearly proportional to the electrical current,  it is also modulated at $\omega_L$ (Fig. \ref{fig:fig3}(a)). This magnetic  field $H^{(CI)}$ further tilts the magnetization, and since this additional tilt is in phase with the magnetization precession, it increases the precession angle. Hence, the magnetization precession itself generates an oscillating magnetic field, amplifying the precession angle within each precession period.  It means that the magnetization precession itself creates an oscillating magnetic field, which enlarges the precision angle within each precession period. 

The enhancement of magnetization precession  can be viewed as a parametric torque induced by the applied DC voltage. When a DC voltage is applied to the MTJ, a small thermal fluctuation seeds the feedback loop and initiates the magnetization precession to grow under the parametric torque until the magnetization reversal occurs.

The parametric torque is calculated as follows. Given that the "pin" layer's magnetization is fixed along the x-axis, the MTJ resistance is determined solely by the x-component of the "free" layer's magnetization. At the precession angle $\theta$, the oscillation of the x-component of the "free" layer's magnetization can be represented as follows: 

\begin{equation}
	{M_x} = {M_{free}} \cdot sin\left( \theta  \right) \cdot sin\left( {\omega t} \right)
	\label{Mx}
\end{equation}

The MTJ resistance is modulated with the precession as:

\begin{equation}
	\begin{array}{l}
		R = {R_0} + \left( {{R_{ \uparrow  \downarrow }} - {R_{ \uparrow  \uparrow }}} \right)\sin \left( \theta  \right) \cdot \sin \left( {\omega t} \right) = \\
		= {R_0}\left[ {1 - {k_{MR}}\sin \left( \theta  \right) \cdot \sin \left( {\omega t} \right)} \right]
	\end{array}
	\label{R_MTJ}
\end{equation}

where $R_0,R_{ \uparrow  \downarrow }, R_{ \uparrow  \uparrow } $  are the MTJ resistance, when the “free” layer is perpendicular  ($\theta=\pi /2$), antiparallel ($\theta=\pi$) and parallel  ($\theta=0$) to the “pin” layer; ${k_{MR}} = \left( {{R_{ \uparrow  \downarrow }} - {R_{ \uparrow  \uparrow }}} \right)/{R_0}$  is the magneto resistance. When a DC voltage $U_{DC}$ is applied to the MTJ, the current is modulated by the resistance oscillation  as

\begin{equation}
	I = \frac{{{U_{DC}}}}{R} = 
	\frac{{{I_{DC}}}}{{1 - {k_{MR}}\sin \left( \theta  \right) \cdot \sin \left( {\omega t} \right)}}
	\label{I_MTJ}
\end{equation}

The experimental measurements conducted in Section 2 confirm that the magnetic field $H^{(CI)}$, resulting from the spin accumulation, is directly proportional to the current $I$. From Eq. \ref{I_MTJ}, $H^{(CI)}$ can be represented as follows:

\begin{equation}
	{H^{(CI)}} = \varsigma I = 
	\frac{{\varsigma  \cdot {I_{DC}}}}{{1 - {k_{MR}}
			\sin \left( \theta  \right) \cdot \sin \left( {\omega t} \right)}}
	\label{H_CI}
\end{equation}

where $\varsigma$  is  the efficiency coefficient for generating $H^{(CI)}$, which has been determined from the measurements in Fig.\ref{fig:fig2}  to be about 1.6 G/(mA/$\mu m^2$) and can be adjusted by an external magnetic field.

By substituting the oscillating magnetic field of Eq. \ref{H_CI} into the Landau-Lifshitz equation and solving it (See Appendix 1), the resulting expression for the parametric torque is as follows: 

\begin{equation}
	{\left( {\frac{{\partial \theta }}{{\partial t}}} \right)_{param}}=
	{\gamma \cdot	\varsigma \cdot  I_{DC} \cdot  k_{MR} \over 2 } \cdot \sin(\theta)
	\label{torque}
\end{equation}

where $\gamma$ is the electron gyromagnetic ratio,.

The parametric torque is not present at equilibrium ($\theta = 0$). Therefore, an initial thermally-activated precession is necessary for the parametric magnetization reversal to occur. After this initial precession, the parametric torque increases as the precession angle $\theta$ increases. For magnetization reversal to happen, the precession torque must be greater than the damping torque at any precession angle $\theta$:
\begin{equation}
	{\left( {\frac{{\partial \theta }}{{\partial t}}} \right)_{param}} > {\left( {\frac{{\partial \theta }}{{\partial t}}} \right)_{damp}} 
	\label{reversal}
\end{equation} 

\section{Discussion and Conclusion}

Similar to the Spin-Transfer Torque (STT)\cite{STT_Torque_Slonczewski_1996,STT_Torque_Berger_1996} and Spin Orbit Torque (SOT)\cite{SOT_Miron2011} mechanisms, the parametric torque can also achieve magnetization reversal of a nanomagnet using a DC electrical current, making it a viable recording mechanism for MRAM\cite{MMM2021_2param,TMRC2021_2param,MMM2020_Hani_Hoff}.

The STT and SOT mechanisms rely on injecting spin-polarized electrons into the nanomagnet to induce magnetization precession and reversal. However, there exists a minimum amount of injected spin-polarized electrons necessary for the magnetization reversal\cite{ZayetsArch2021Parametric}, which imposes a limitation on the critical current required for the recording. This limitation results in a relatively high minimal recording current, thereby affecting the performance of MRAM.

In contrast, the parametric magnetization reversal operates on a resonance principle, enabling it to overcome the restrictions imposed by the minimum required amount of injected spin-polarized electrons. Consequently, it has the potential to be optimized for a lower recording current, offering improved performance for MRAM.

The parametric magnetization reversal is driven by the magnetic field created by spin- accumulated electrons. When an electrical current passes through a ferromagnetic nanomagnet, the Spin Hall effect\cite{SpinHallKato2004,SpinHallDyakonov1971} and the Ordinary Hall effect\cite{ZayetsJMMM2018Holes} cause spin accumulation at its boundaries. As a result, any magnetic structure with magneto-resistance (MR) has the potential to experience parametric torque. This suggests that the parametric torque, along with STT or SOT, may contribute to the magnetization reversal process in already- existing MRAM systems.

Depending on the circumstances, the parametric torque can have the same polarity as STT or SOT, leading to improved recording performance. Conversely, it may have an opposite polarity to STT or SOT, impeding the magnetization reversal process. The parametric torque is highly specific to each MRAM cell's structure and design, and understanding and optimizing it is crucial to enhance the reliability and efficiency of modern MRAM devices. Identifying and fine-tuning the parametric torque can be a valuable approach to achieve more advanced and dependable MRAM operations.

The existence of magneto-resistance (MR) alone is not sufficient for the presence of parametric torque; the modulation of the electrical current by the magnetization precession is essential. When the magnetization precession axis aligns with the easy magnetic axis of the ``pin'' layer, there is no precession modulation of the current. In such cases, during precession, the angle between the magnetizations of the ``free'' and ``pin'' layers remains unchanged, and the resistance is not modulated. This alignment can occur when the easy magnetic axes of the ``free'' and ``pin'' layers are oriented in the same direction. It is  a common configuration for conventional MRAM that allows for maximum change in resistance between two opposite equilibrium directions of the ``free'' layer, thus facilitating memory reading. Hence, it may be presumed that the parametric torque is inconsequential for the typical configuration of MRAM. However, this assumption is incorrect.

Even in cases of parallel alignment, which is unfavorable  for parametric torque, there is still some precession modulation of the current and, consequently, some parametric torque. In the case of in-plane parallel alignment of easy axes, the precession axis is tilted from the in-plane direction due to the anisotropy caused by the finite thickness (Kittel effect\cite{FMR_Kittel}), contributing to the parametric torque. In situations where the alignment is perpendicular to the plane, the axes are misaligned due to the current-independent component of the magnetic field observed in Fig.1. Additionally, minor fabrication imperfections may also lead to some axis misalignment.

The external magnetic field may  play a critical role in the magnetization reversal, especially in cases where the equilibrium magnetizations of the "free" and "pin" layers are aligned in parallel. In such scenarios, the external magnetic field causes a misalignment in magnetization, significantly enhancing the parametric mechanism of magnetization reversal. Without this external influence, magnetic reversal might be unachievable for a particular sample, even in the presence of STT or SOT.  It's solely the inclusion of the parametric mechanism that triggers magnetization reversal. Thus, the presence of an external magnetic field becomes crucial to enable magnetization reversal in such specific cases. An in-plane external magnetic field is necessary and often unavoidable for magnetization reversal by Spin Orbit Torque (SOT)\cite{SOT_Miron2011} or the magnetization reversal by a gate voltage using the Voltage-Controlled Magnetic Anisotropy (VCMA) effect\cite{VCMA_Shiota2012,VCMA_Nozaki2016}. The necessity of an in-plane external field in the magnetization reversal process suggests a significant contribution of the parametric torque in the overall reversal mechanism.

A compelling indicator of the presence of parametric torque in a specific MTJ is its use as a Spin-torque oscillator (STO)\cite{STO.PRL.1st,STO.2004.PRB.2nd}. In an STO, a RF current is generated by the modulation of MTJ resistance due to precession, leading to the excitation of microwaves utilized as a microwave source in RF circuits\cite{STO.NIST.2005,STO.AIST.FERT.2010.basic}. The modulation of current driven by magnetization precession plays a crucial role in STOs. As previously mentioned, electrical currents typically create spin accumulation in nanomagnets. The modulated current and the resulting modulated spin accumulation create the conditions necessary for the positive feedback loop required for the existence of parametric torque. Therefore, the conditions that enable a MTJ to function as a Spin-torque oscillator are similar to the conditions that support the existence of the parametric torque.

Indeed, the presence of the positive feedback loop, depicted in Fig.4, was both theoretically studied and experimentally verified in the Spin-torque oscillator  using an external feedback loop and bulk microwave components\cite{STO.AIST.2016.feedback.magnetic.field,STO.AIST.2016.delayFeedback,STO.AIST.2018.feedback.magnetic.field}. In this setup, the microwave emitted by the STO was coupled into a microwave waveguide. A section of the waveguide was positioned in close proximity to the MTJ, enabling the magnetic field of the waveguide mode to penetrate the MTJ and tilt the magnetization of the "free" layer. The frequency and phase of the microwave, and thus the magnetization tilt, were precisely synchronized with the magnetization precession, creating a positive feedback loop. To further enhance the performance of the STO, a microwave amplifier and phase shifter were incorporated into the feedback loop. The utilization of this external feedback loop substantially improved the quality factor of the STO, surpassing 10,000\cite{STO.AIST.2016.feedback.magnetic.field}. Despite the use of external bulk microwave components, this setup demonstrated the capability to achieve parametric amplification of magnetization precession in a MTJ, indicating the potential for creating a substantial parametric torque.

In addition to its resonance nature, there is another significant difference between the described parametric mechanism of magnetization reversal and the conventional Spin-Transfer Torque (STT) and Spin-Orbit Torque (SOT).  In the case of the STT or SOT torque, spin-polarized conduction electrons diffuse into the bulk of the nanomagnet, interacting with the localized d-electrons and transferring their spin to them. In contrast, in the described mechanism, the magnetic field created by the spin-polarized conduction electrons interacts with the spins of the localized d-electrons. Spin diffusion is not required for this mechanism, and the spin accumulation can be even spatially separated from the localized d-electrons of the nanomagnet.

The characteristics and attributes of the magnetic field generated by spin accumulation must be examined in conjunction with the behavior of the spin accumulation and the diverse dynamics of spin-polarized conduction electrons. The properties of this magnetic field are inherently linked to those of the spin accumulation. Hence, comprehending the behavior of the magnetic field induced by spin accumulation and the torque it generates is intricately interrelated with the properties and dynamics of Spin-Transfer Torque  and Spin-Orbit Torque.

In conclusion, the parametric torque presents a promising approach for achieving magnetization reversal in nanomagnets using a DC electrical current. This resonance torque, in conjunction with Spin-Transfer Torque and Spin Orbit Torque, can serve as an efficient recording mechanism in MRAM. The key components contributing to this torque are the magnetic field of spin accumulation in a nanomagnet and the modulation of current by magnetization precession, which together create a positive feedback loop. This feedback loop leads to enhanced precession and ultimately facilitates the magnetization reversal process. With its potential for low-power and high-performance operation, the parametric torque holds considerable promise for advancing modern MRAM technology.


\bibliography{bibParametricReversal}

\providecommand{\noopsort}[1]{}\providecommand{\singleletter}[1]{#1}%
\begin{thebibliography}{40}%
\makeatletter
\providecommand \@ifxundefined [1]{%
 \@ifx{#1\undefined}
}%
\providecommand \@ifnum [1]{%
 \ifnum #1\expandafter \@firstoftwo
 \else \expandafter \@secondoftwo
 \fi
}%
\providecommand \@ifx [1]{%
 \ifx #1\expandafter \@firstoftwo
 \else \expandafter \@secondoftwo
 \fi
}%
\providecommand \natexlab [1]{#1}%
\providecommand \enquote  [1]{``#1''}%
\providecommand \bibnamefont  [1]{#1}%
\providecommand \bibfnamefont [1]{#1}%
\providecommand \citenamefont [1]{#1}%
\providecommand \href@noop [0]{\@secondoftwo}%
\providecommand \href [0]{\begingroup \@sanitize@url \@href}%
\providecommand \@href[1]{\@@startlink{#1}\@@href}%
\providecommand \@@href[1]{\endgroup#1\@@endlink}%
\providecommand \@sanitize@url [0]{\catcode `\\12\catcode `\$12\catcode
  `\&12\catcode `\#12\catcode `\^12\catcode `\_12\catcode `\%12\relax}%
\providecommand \@@startlink[1]{}%
\providecommand \@@endlink[0]{}%
\providecommand \url  [0]{\begingroup\@sanitize@url \@url }%
\providecommand \@url [1]{\endgroup\@href {#1}{\urlprefix }}%
\providecommand \urlprefix  [0]{URL }%
\providecommand \Eprint [0]{\href }%
\providecommand \doibase [0]{https://doi.org/}%
\providecommand \selectlanguage [0]{\@gobble}%
\providecommand \bibinfo  [0]{\@secondoftwo}%
\providecommand \bibfield  [0]{\@secondoftwo}%
\providecommand \translation [1]{[#1]}%
\providecommand \BibitemOpen [0]{}%
\providecommand \bibitemStop [0]{}%
\providecommand \bibitemNoStop [0]{.\EOS\space}%
\providecommand \EOS [0]{\spacefactor3000\relax}%
\providecommand \BibitemShut  [1]{\csname bibitem#1\endcsname}%
\let\auto@bib@innerbib\@empty
\bibitem [{\citenamefont {Cubukcu}\ \emph {et~al.}(2018)\citenamefont
  {Cubukcu}, \citenamefont {Boulle}, \citenamefont {Mikuszeit}, \citenamefont
  {Hamelin}, \citenamefont {Brächer}, \citenamefont {Lamard}, \citenamefont
  {Cyrille}, \citenamefont {Buda-Prejbeanu}, \citenamefont {Garello},
  \citenamefont {Miron}, \citenamefont {Klein}, \citenamefont {de~Loubens},
  \citenamefont {Naletov}, \citenamefont {Langer}, \citenamefont {Ocker},
  \citenamefont {Gambardella},\ and\ \citenamefont {Gaudin}}]{MRAMCubukcu}%
  \BibitemOpen
  \bibfield  {author} {\bibinfo {author} {\bibfnamefont {M.}~\bibnamefont
  {Cubukcu}}, \bibinfo {author} {\bibfnamefont {O.}~\bibnamefont {Boulle}},
  \bibinfo {author} {\bibfnamefont {N.}~\bibnamefont {Mikuszeit}}, \bibinfo
  {author} {\bibfnamefont {C.}~\bibnamefont {Hamelin}}, \bibinfo {author}
  {\bibfnamefont {T.}~\bibnamefont {Brächer}}, \bibinfo {author}
  {\bibfnamefont {N.}~\bibnamefont {Lamard}}, \bibinfo {author} {\bibfnamefont
  {M.-C.}\ \bibnamefont {Cyrille}}, \bibinfo {author} {\bibfnamefont
  {L.}~\bibnamefont {Buda-Prejbeanu}}, \bibinfo {author} {\bibfnamefont
  {K.}~\bibnamefont {Garello}}, \bibinfo {author} {\bibfnamefont {I.~M.}\
  \bibnamefont {Miron}}, \bibinfo {author} {\bibfnamefont {O.}~\bibnamefont
  {Klein}}, \bibinfo {author} {\bibfnamefont {G.}~\bibnamefont {de~Loubens}},
  \bibinfo {author} {\bibfnamefont {V.~V.}\ \bibnamefont {Naletov}}, \bibinfo
  {author} {\bibfnamefont {J.}~\bibnamefont {Langer}}, \bibinfo {author}
  {\bibfnamefont {B.}~\bibnamefont {Ocker}}, \bibinfo {author} {\bibfnamefont
  {P.}~\bibnamefont {Gambardella}},\ and\ \bibinfo {author} {\bibfnamefont
  {G.}~\bibnamefont {Gaudin}},\ }\bibfield  {title} {\bibinfo {title}
  {Ultra-fast perpendicular spin–orbit torque mram},\ }\href
  {https://doi.org/10.1109/TMAG.2017.2772185} {\bibfield  {journal} {\bibinfo
  {journal} {IEEE Transactions on Magnetics}\ }\textbf {\bibinfo {volume}
  {54}},\ \bibinfo {pages} {1} (\bibinfo {year} {2018})}\BibitemShut {NoStop}%
\bibitem [{\citenamefont {Valet}\ and\ \citenamefont {Fert}(1993)}]{FertpGMR}%
  \BibitemOpen
  \bibfield  {author} {\bibinfo {author} {\bibfnamefont {T.}~\bibnamefont
  {Valet}}\ and\ \bibinfo {author} {\bibfnamefont {A.}~\bibnamefont {Fert}},\
  }\bibfield  {title} {\bibinfo {title} {Theory of the perpendicular
  magnetoresistance in magnetic multilayers},\ }\href
  {https://doi.org/10.1103/PhysRevB.48.7099} {\bibfield  {journal} {\bibinfo
  {journal} {Phys. Rev. B}\ }\textbf {\bibinfo {volume} {48}},\ \bibinfo
  {pages} {7099} (\bibinfo {year} {1993})}\BibitemShut {NoStop}%
\bibitem [{\citenamefont {Flederling}\ \emph {et~al.}(1999)\citenamefont
  {Flederling}, \citenamefont {Kelm}, \citenamefont {Reuscher}, \citenamefont
  {Ossau}, \citenamefont {Schmidt}, \citenamefont {Waag},\ and\ \citenamefont
  {Molenkamp}}]{Flederling1999SpinInjectionLD}%
  \BibitemOpen
  \bibfield  {author} {\bibinfo {author} {\bibfnamefont {R.}~\bibnamefont
  {Flederling}}, \bibinfo {author} {\bibfnamefont {M.}~\bibnamefont {Kelm}},
  \bibinfo {author} {\bibfnamefont {G.}~\bibnamefont {Reuscher}}, \bibinfo
  {author} {\bibfnamefont {W.}~\bibnamefont {Ossau}}, \bibinfo {author}
  {\bibfnamefont {G.}~\bibnamefont {Schmidt}}, \bibinfo {author} {\bibfnamefont
  {A.}~\bibnamefont {Waag}},\ and\ \bibinfo {author} {\bibfnamefont {L.~W.}\
  \bibnamefont {Molenkamp}},\ }\bibfield  {title} {\bibinfo {title} {{Injection
  and detection of a spin-polarized current in a light-emitting diode}},\
  }\href {https://doi.org/10.1038/45502} {\bibfield  {journal} {\bibinfo
  {journal} {Nature}\ }\textbf {\bibinfo {volume} {402}},\ \bibinfo {pages}
  {787} (\bibinfo {year} {1999})}\BibitemShut {NoStop}%
\bibitem [{\citenamefont {Crooker}\ and\ \citenamefont
  {Smith}(2005)}]{Crooker2005SpinInjectionImage}%
  \BibitemOpen
  \bibfield  {author} {\bibinfo {author} {\bibfnamefont {S.~A.}\ \bibnamefont
  {Crooker}}\ and\ \bibinfo {author} {\bibfnamefont {D.~L.}\ \bibnamefont
  {Smith}},\ }\bibfield  {title} {\bibinfo {title} {Imaging spin flows in
  semiconductors subject to electric, magnetic, and strain fields},\ }\href
  {https://doi.org/10.1103/PhysRevLett.94.236601} {\bibfield  {journal}
  {\bibinfo  {journal} {Phys. Rev. Lett.}\ }\textbf {\bibinfo {volume} {94}},\
  \bibinfo {pages} {236601} (\bibinfo {year} {2005})}\BibitemShut {NoStop}%
\bibitem [{\citenamefont {Ohno}\ \emph {et~al.}(1999)\citenamefont {Ohno},
  \citenamefont {Young}, \citenamefont {Beschoten}, \citenamefont {Matsukura},
  \citenamefont {Ohno},\ and\ \citenamefont
  {Awschalom}}]{Ohno1999_SpinInjection}%
  \BibitemOpen
  \bibfield  {author} {\bibinfo {author} {\bibfnamefont {Y.}~\bibnamefont
  {Ohno}}, \bibinfo {author} {\bibfnamefont {D.~K.}\ \bibnamefont {Young}},
  \bibinfo {author} {\bibfnamefont {B.}~\bibnamefont {Beschoten}}, \bibinfo
  {author} {\bibfnamefont {F.}~\bibnamefont {Matsukura}}, \bibinfo {author}
  {\bibfnamefont {H.}~\bibnamefont {Ohno}},\ and\ \bibinfo {author}
  {\bibfnamefont {D.~D.}\ \bibnamefont {Awschalom}},\ }\bibfield  {title}
  {\bibinfo {title} {{Electrical spin injection in a ferromagnetic
  semiconductor heterostructure}},\ }\href {https://doi.org/10.1038/45509}
  {\bibfield  {journal} {\bibinfo  {journal} {Nature}\ }\textbf {\bibinfo
  {volume} {402}},\ \bibinfo {pages} {790} (\bibinfo {year}
  {1999})}\BibitemShut {NoStop}%
\bibitem [{\citenamefont {Crooker}\ \emph {et~al.}(2005)\citenamefont
  {Crooker}, \citenamefont {Furis}, \citenamefont {Lou}, \citenamefont
  {Adelmann}, \citenamefont {Smith}, \citenamefont {Palmstr{\o}m},\ and\
  \citenamefont {Crowell}}]{Crooker2005bImagingSpinTransport}%
  \BibitemOpen
  \bibfield  {author} {\bibinfo {author} {\bibfnamefont {S.~A.}\ \bibnamefont
  {Crooker}}, \bibinfo {author} {\bibfnamefont {M.}~\bibnamefont {Furis}},
  \bibinfo {author} {\bibfnamefont {X.}~\bibnamefont {Lou}}, \bibinfo {author}
  {\bibfnamefont {C.}~\bibnamefont {Adelmann}}, \bibinfo {author}
  {\bibfnamefont {D.~L.}\ \bibnamefont {Smith}}, \bibinfo {author}
  {\bibfnamefont {C.~J.}\ \bibnamefont {Palmstr{\o}m}},\ and\ \bibinfo {author}
  {\bibfnamefont {P.~A.}\ \bibnamefont {Crowell}},\ }\bibfield  {title}
  {\bibinfo {title} {{Applied physics: Imaging spin transport in lateral
  ferromagnet/ semiconductor structures}},\ }\href
  {https://doi.org/10.1126/science.1116865} {\bibfield  {journal} {\bibinfo
  {journal} {Science}\ }\textbf {\bibinfo {volume} {309}},\ \bibinfo {pages}
  {2191} (\bibinfo {year} {2005})}\BibitemShut {NoStop}%
\bibitem [{\citenamefont
  {Zayets}(2021{\natexlab{a}})}]{ZayetsArch2021Parametric}%
  \BibitemOpen
  \bibfield  {author} {\bibinfo {author} {\bibfnamefont {V.}~\bibnamefont
  {Zayets}},\ }\bibfield  {title} {\bibinfo {title} {Mechanism of parametric
  pumping of magnetization precession in a nanomagnet. parametric mechanism of
  current-induced magnetization reversal},\ }\bibfield  {journal} {\bibinfo
  {journal} {arXiv}\ }\textbf {\bibinfo {volume} {2104}},\ \href
  {https://doi.org/10.48550/arxiv.2104.13008} {10.48550/arxiv.2104.13008}
  (\bibinfo {year} {2021}{\natexlab{a}})\BibitemShut {NoStop}%
\bibitem [{\citenamefont {Kittel}(1948)}]{FMR_Kittel}%
  \BibitemOpen
  \bibfield  {author} {\bibinfo {author} {\bibfnamefont {C.}~\bibnamefont
  {Kittel}},\ }\bibfield  {title} {\bibinfo {title} {On the theory of
  ferromagnetic resonance absorption},\ }\href
  {https://doi.org/10.1103/PhysRev.73.155} {\bibfield  {journal} {\bibinfo
  {journal} {Phys. Rev.}\ }\textbf {\bibinfo {volume} {73}},\ \bibinfo {pages}
  {155} (\bibinfo {year} {1948})}\BibitemShut {NoStop}%
\bibitem [{\citenamefont {Zayets}(2022)}]{Zayets2022IEEE_Hci}%
  \BibitemOpen
  \bibfield  {author} {\bibinfo {author} {\bibfnamefont {V.}~\bibnamefont
  {Zayets}},\ }\bibfield  {title} {\bibinfo {title} {Measurement of magnetic
  field induced by spin-accumulated electrons in a fecob nanomagnet},\ }\href
  {https://doi.org/10.1109/TMAG.2021.3111598} {\bibfield  {journal} {\bibinfo
  {journal} {IEEE Transactions on Magnetics}\ }\textbf {\bibinfo {volume}
  {58}},\ \bibinfo {pages} {1} (\bibinfo {year} {2022})}\BibitemShut {NoStop}%
\bibitem [{\citenamefont {Garello}\ \emph {et~al.}(2013)\citenamefont
  {Garello}, \citenamefont {Miron}, \citenamefont {Avci}, \citenamefont
  {Freimuth}, \citenamefont {Mokrousov}, \citenamefont {Bl{\"{u}}gel},
  \citenamefont {Auffret}, \citenamefont {Boulle}, \citenamefont {Gaudin},\
  and\ \citenamefont {Gambardella}}]{SecondHarmonic_Garello2013}%
  \BibitemOpen
  \bibfield  {author} {\bibinfo {author} {\bibfnamefont {K.}~\bibnamefont
  {Garello}}, \bibinfo {author} {\bibfnamefont {I.~M.}\ \bibnamefont {Miron}},
  \bibinfo {author} {\bibfnamefont {C.~O.}\ \bibnamefont {Avci}}, \bibinfo
  {author} {\bibfnamefont {F.}~\bibnamefont {Freimuth}}, \bibinfo {author}
  {\bibfnamefont {Y.}~\bibnamefont {Mokrousov}}, \bibinfo {author}
  {\bibfnamefont {S.}~\bibnamefont {Bl{\"{u}}gel}}, \bibinfo {author}
  {\bibfnamefont {S.}~\bibnamefont {Auffret}}, \bibinfo {author} {\bibfnamefont
  {O.}~\bibnamefont {Boulle}}, \bibinfo {author} {\bibfnamefont
  {G.}~\bibnamefont {Gaudin}},\ and\ \bibinfo {author} {\bibfnamefont
  {P.}~\bibnamefont {Gambardella}},\ }\bibfield  {title} {\bibinfo {title}
  {{Suppl Symmetry and magnitude of spin-orbit torques in ferromagnetic
  heterostructures}},\ }\href {https://doi.org/10.1038/nnano.2013.145}
  {\bibfield  {journal} {\bibinfo  {journal} {Nature Nanotechnology}\ }\textbf
  {\bibinfo {volume} {8}},\ \bibinfo {pages} {587} (\bibinfo {year}
  {2013})}\BibitemShut {NoStop}%
\bibitem [{\citenamefont {Kim}\ \emph {et~al.}(2013)\citenamefont {Kim},
  \citenamefont {Sinha}, \citenamefont {Hayashi}, \citenamefont {Yamanouchi},
  \citenamefont {Fukami}, \citenamefont {Suzuki}, \citenamefont {Mitani},\ and\
  \citenamefont {Ohno}}]{SecondHarmonic_Kim2013}%
  \BibitemOpen
  \bibfield  {author} {\bibinfo {author} {\bibfnamefont {J.}~\bibnamefont
  {Kim}}, \bibinfo {author} {\bibfnamefont {J.}~\bibnamefont {Sinha}}, \bibinfo
  {author} {\bibfnamefont {M.}~\bibnamefont {Hayashi}}, \bibinfo {author}
  {\bibfnamefont {M.}~\bibnamefont {Yamanouchi}}, \bibinfo {author}
  {\bibfnamefont {S.}~\bibnamefont {Fukami}}, \bibinfo {author} {\bibfnamefont
  {T.}~\bibnamefont {Suzuki}}, \bibinfo {author} {\bibfnamefont
  {S.}~\bibnamefont {Mitani}},\ and\ \bibinfo {author} {\bibfnamefont
  {H.}~\bibnamefont {Ohno}},\ }\bibfield  {title} {\bibinfo {title} {{Layer
  thickness dependence of the current-induced effective field vector in
  Ta|CoFeB|MgO}},\ }\href {https://doi.org/10.1038/nmat3522} {\bibfield
  {journal} {\bibinfo  {journal} {Nature Materials}\ }\textbf {\bibinfo
  {volume} {12}},\ \bibinfo {pages} {240} (\bibinfo {year} {2013})}\BibitemShut
  {NoStop}%
\bibitem [{\citenamefont {Abert}\ \emph {et~al.}(2015)\citenamefont {Abert},
  \citenamefont {Ruggeri}, \citenamefont {Bruckner}, \citenamefont {Vogler},
  \citenamefont {Hrkac}, \citenamefont {Praetorius},\ and\ \citenamefont
  {Suess}}]{FielLikeTorqueViennaTheory2015}%
  \BibitemOpen
  \bibfield  {author} {\bibinfo {author} {\bibfnamefont {C.}~\bibnamefont
  {Abert}}, \bibinfo {author} {\bibfnamefont {M.}~\bibnamefont {Ruggeri}},
  \bibinfo {author} {\bibfnamefont {F.}~\bibnamefont {Bruckner}}, \bibinfo
  {author} {\bibfnamefont {C.}~\bibnamefont {Vogler}}, \bibinfo {author}
  {\bibfnamefont {G.}~\bibnamefont {Hrkac}}, \bibinfo {author} {\bibfnamefont
  {D.}~\bibnamefont {Praetorius}},\ and\ \bibinfo {author} {\bibfnamefont
  {D.}~\bibnamefont {Suess}},\ }\bibfield  {title} {\bibinfo {title} {A
  three-dimensional spin-diffusion model for micromagnetics},\ }\href
  {https://doi.org/10.1038/srep14855} {\bibfield  {journal} {\bibinfo
  {journal} {Scientific Reports}\ }\textbf {\bibinfo {volume} {5}},\ \bibinfo
  {pages} {2045} (\bibinfo {year} {2015})}\BibitemShut {NoStop}%
\bibitem [{\citenamefont {Lepadatu}(2017)}]{FielLikeTorqueUKTheory2017}%
  \BibitemOpen
  \bibfield  {author} {\bibinfo {author} {\bibfnamefont {S.}~\bibnamefont
  {Lepadatu}},\ }\bibfield  {title} {\bibinfo {title} {Unified treatment of
  spin torques using a coupled magnetisation dynamics and three-dimensional
  spin current solver},\ }\href {https://doi.org/10.1038/s41598-017-13181-x}
  {\bibfield  {journal} {\bibinfo  {journal} {Scientific Reports}\ }\textbf
  {\bibinfo {volume} {7}},\ \bibinfo {pages} {12937} (\bibinfo {year}
  {2017})}\BibitemShut {NoStop}%
\bibitem [{\citenamefont {Legrand}\ \emph {et~al.}(2015)\citenamefont
  {Legrand}, \citenamefont {Ramaswamy}, \citenamefont {Mishra},\ and\
  \citenamefont {Yang}}]{FielLikeTorqueTheory2015}%
  \BibitemOpen
  \bibfield  {author} {\bibinfo {author} {\bibfnamefont {W.}~\bibnamefont
  {Legrand}}, \bibinfo {author} {\bibfnamefont {R.}~\bibnamefont {Ramaswamy}},
  \bibinfo {author} {\bibfnamefont {R.}~\bibnamefont {Mishra}},\ and\ \bibinfo
  {author} {\bibfnamefont {H.}~\bibnamefont {Yang}},\ }\bibfield  {title}
  {\bibinfo {title} {Coherent subnanosecond switching of perpendicular
  magnetization by the fieldlike spin-orbit torque without an external magnetic
  field},\ }\href
  {https://doi.org/10.1103/PHYSREVAPPLIED.3.064012/FIGURES/6/MEDIUM} {\bibfield
   {journal} {\bibinfo  {journal} {Physical Review Applied}\ }\textbf {\bibinfo
  {volume} {3}},\ \bibinfo {pages} {064012} (\bibinfo {year}
  {2015})}\BibitemShut {NoStop}%
\bibitem [{\citenamefont {Pi}\ \emph {et~al.}(2010)\citenamefont {Pi},
  \citenamefont {Kim}, \citenamefont {Bae}, \citenamefont {Lee}, \citenamefont
  {Cho}, \citenamefont {Kim},\ and\ \citenamefont
  {Seo}}]{SecondHarmonic_Pi2010}%
  \BibitemOpen
  \bibfield  {author} {\bibinfo {author} {\bibfnamefont {U.~H.}\ \bibnamefont
  {Pi}}, \bibinfo {author} {\bibfnamefont {K.~W.}\ \bibnamefont {Kim}},
  \bibinfo {author} {\bibfnamefont {J.~Y.}\ \bibnamefont {Bae}}, \bibinfo
  {author} {\bibfnamefont {S.~C.}\ \bibnamefont {Lee}}, \bibinfo {author}
  {\bibfnamefont {Y.~J.}\ \bibnamefont {Cho}}, \bibinfo {author} {\bibfnamefont
  {K.~S.}\ \bibnamefont {Kim}},\ and\ \bibinfo {author} {\bibfnamefont
  {S.}~\bibnamefont {Seo}},\ }\bibfield  {title} {\bibinfo {title} {{Tilting of
  the spin orientation induced by Rashba effect in ferromagnetic metal
  layer}},\ }\href {https://doi.org/10.1063/1.3502596} {\bibfield  {journal}
  {\bibinfo  {journal} {Applied Physics Letters}\ }\textbf {\bibinfo {volume}
  {97}},\ \bibinfo {pages} {162507} (\bibinfo {year} {2010})}\BibitemShut
  {NoStop}%
\bibitem [{\citenamefont {Zayets}(2020)}]{MMM2020_Hani_Hoff}%
  \BibitemOpen
  \bibfield  {author} {\bibinfo {author} {\bibfnamefont {V.}~\bibnamefont
  {Zayets}},\ }\bibfield  {title} {\bibinfo {title} {Measurement of anisotropy
  field under external perpendicular magnetic field in fecob and feb
  nanomagnets. study of pma features in a nanomagnet},\ }in\ \href@noop {}
  {\emph {\bibinfo {booktitle} {MMM 2020.The 65th Annual Conference on
  Magnetism and Magnetic Materials}}}\ (\bibinfo {year} {2020})\ pp.\ \bibinfo
  {pages} {N4--03}\BibitemShut {NoStop}%
\bibitem [{\citenamefont {Zayets}(2021{\natexlab{b}})}]{TMRC2021_2param}%
  \BibitemOpen
  \bibfield  {author} {\bibinfo {author} {\bibfnamefont {V.}~\bibnamefont
  {Zayets}},\ }\bibfield  {title} {\bibinfo {title} {Comparison of two
  parametric mechanisms of magnetization reversal in fecob nanomagnet. theory
  and experiment},\ }in\ \href@noop {} {\emph {\bibinfo {booktitle} {TMRC
  2021.The 32nd Magnetic Recording Conference}}}\ (\bibinfo {year} {2021})\
  p.~\bibinfo {pages} {H4}\BibitemShut {NoStop}%
\bibitem [{\citenamefont {Zayets}(2021{\natexlab{c}})}]{MMM2021_2param}%
  \BibitemOpen
  \bibfield  {author} {\bibinfo {author} {\bibfnamefont {V.}~\bibnamefont
  {Zayets}},\ }\bibfield  {title} {\bibinfo {title} {Experimental evaluation of
  two parametric mechanisms of magnetization reversal in fecob nanomagnet},\
  }in\ \href@noop {} {\emph {\bibinfo {booktitle} {MMM 2021.The 66th Annual
  Conference on Magnetism and Magnetic Materials}}}\ (\bibinfo {year} {2021})\
  pp.\ \bibinfo {pages} {IOA--07}\BibitemShut {NoStop}%
\bibitem [{\citenamefont {Zayets}\ and\ \citenamefont
  {Mishchenko}(2020{\natexlab{a}})}]{MMM2020_sp}%
  \BibitemOpen
  \bibfield  {author} {\bibinfo {author} {\bibfnamefont {V.}~\bibnamefont
  {Zayets}}\ and\ \bibinfo {author} {\bibfnamefont {A.~S.}\ \bibnamefont
  {Mishchenko}},\ }\bibfield  {title} {\bibinfo {title} {Study of magnetic
  field dependence of hall effect in feb and fecob nanomagnets. an evidence of
  contribution of inverse spin hall effect},\ }in\ \href@noop {} {\emph
  {\bibinfo {booktitle} {MMM 2020.The 65th Annual Conference on Magnetism and
  Magnetic Materials}}}\ (\bibinfo {year} {2020})\ pp.\ \bibinfo {pages}
  {P6--07}\BibitemShut {NoStop}%
\bibitem [{\citenamefont {Zayets}\ and\ \citenamefont
  {Mishchenko}(2020{\natexlab{b}})}]{Zayets2020MishenkoSpinPol}%
  \BibitemOpen
  \bibfield  {author} {\bibinfo {author} {\bibfnamefont {V.}~\bibnamefont
  {Zayets}}\ and\ \bibinfo {author} {\bibfnamefont {A.~S.}\ \bibnamefont
  {Mishchenko}},\ }\bibfield  {title} {\bibinfo {title} {Hall effect in
  ferromagnetic nanomagnets: Magnetic field dependence as evidence of inverse
  spin hall effect contribution},\ }\href
  {https://doi.org/10.1103/PhysRevB.102.100404} {\bibfield  {journal} {\bibinfo
   {journal} {Phys. Rev. B}\ }\textbf {\bibinfo {volume} {102}},\ \bibinfo
  {pages} {100404} (\bibinfo {year} {2020}{\natexlab{b}})}\BibitemShut
  {NoStop}%
\bibitem [{\citenamefont {Dyakonov}\ and\ \citenamefont
  {Perel}(1971)}]{SpinHallDyakonov1971}%
  \BibitemOpen
  \bibfield  {author} {\bibinfo {author} {\bibfnamefont {M.}~\bibnamefont
  {Dyakonov}}\ and\ \bibinfo {author} {\bibfnamefont {V.}~\bibnamefont
  {Perel}},\ }\bibfield  {title} {\bibinfo {title} {Current-induced spin
  orientation of electrons in semiconductors},\ }\href
  {https://doi.org/https://doi.org/10.1016/0375-9601(71)90196-4} {\bibfield
  {journal} {\bibinfo  {journal} {Physics Letters A}\ }\textbf {\bibinfo
  {volume} {35}},\ \bibinfo {pages} {459} (\bibinfo {year} {1971})}\BibitemShut
  {NoStop}%
\bibitem [{\citenamefont {Kato}\ \emph {et~al.}(2004)\citenamefont {Kato},
  \citenamefont {Myers}, \citenamefont {Gossard},\ and\ \citenamefont
  {Awschalom}}]{SpinHallKato2004}%
  \BibitemOpen
  \bibfield  {author} {\bibinfo {author} {\bibfnamefont {Y.~K.}\ \bibnamefont
  {Kato}}, \bibinfo {author} {\bibfnamefont {R.~C.}\ \bibnamefont {Myers}},
  \bibinfo {author} {\bibfnamefont {A.~C.}\ \bibnamefont {Gossard}},\ and\
  \bibinfo {author} {\bibfnamefont {D.~D.}\ \bibnamefont {Awschalom}},\
  }\bibfield  {title} {\bibinfo {title} {{Observation of the Spin Hall Effect
  in Semiconductors}},\ }\href {https://doi.org/10.1126/science.1105514}
  {\bibfield  {journal} {\bibinfo  {journal} {Science}\ }\textbf {\bibinfo
  {volume} {306}},\ \bibinfo {pages} {1910} (\bibinfo {year}
  {2004})}\BibitemShut {NoStop}%
\bibitem [{\citenamefont {Wunderlich}\ \emph {et~al.}(2005)\citenamefont
  {Wunderlich}, \citenamefont {Kaestner}, \citenamefont {Sinova},\ and\
  \citenamefont {Jungwirth}}]{SpinHallSinolva2005}%
  \BibitemOpen
  \bibfield  {author} {\bibinfo {author} {\bibfnamefont {J.}~\bibnamefont
  {Wunderlich}}, \bibinfo {author} {\bibfnamefont {B.}~\bibnamefont
  {Kaestner}}, \bibinfo {author} {\bibfnamefont {J.}~\bibnamefont {Sinova}},\
  and\ \bibinfo {author} {\bibfnamefont {T.}~\bibnamefont {Jungwirth}},\
  }\bibfield  {title} {\bibinfo {title} {Experimental observation of the
  spin-hall effect in a two-dimensional spin-orbit coupled semiconductor
  system},\ }\href {https://doi.org/10.1103/PhysRevLett.94.047204} {\bibfield
  {journal} {\bibinfo  {journal} {Phys. Rev. Lett.}\ }\textbf {\bibinfo
  {volume} {94}},\ \bibinfo {pages} {047204} (\bibinfo {year}
  {2005})}\BibitemShut {NoStop}%
\bibitem [{\citenamefont {Hirsch}(1999)}]{SpinHallHirsh1999}%
  \BibitemOpen
  \bibfield  {author} {\bibinfo {author} {\bibfnamefont {J.~E.}\ \bibnamefont
  {Hirsch}},\ }\bibfield  {title} {\bibinfo {title} {Spin hall effect},\ }\href
  {https://doi.org/10.1103/PhysRevLett.83.1834} {\bibfield  {journal} {\bibinfo
   {journal} {Phys. Rev. Lett.}\ }\textbf {\bibinfo {volume} {83}},\ \bibinfo
  {pages} {1834} (\bibinfo {year} {1999})}\BibitemShut {NoStop}%
\bibitem [{\citenamefont {Kondou}\ \emph {et~al.}(2021)\citenamefont {Kondou},
  \citenamefont {Chen}, \citenamefont {Tomita}, \citenamefont {Ikhlas},
  \citenamefont {Higo}, \citenamefont {MacDonald}, \citenamefont {Nakatsuji},\
  and\ \citenamefont {Otani}}]{antiferroFMRonCurrent2021}%
  \BibitemOpen
  \bibfield  {author} {\bibinfo {author} {\bibfnamefont {K.}~\bibnamefont
  {Kondou}}, \bibinfo {author} {\bibfnamefont {H.}~\bibnamefont {Chen}},
  \bibinfo {author} {\bibfnamefont {T.}~\bibnamefont {Tomita}}, \bibinfo
  {author} {\bibfnamefont {M.}~\bibnamefont {Ikhlas}}, \bibinfo {author}
  {\bibfnamefont {T.}~\bibnamefont {Higo}}, \bibinfo {author} {\bibfnamefont
  {A.~H.}\ \bibnamefont {MacDonald}}, \bibinfo {author} {\bibfnamefont
  {S.}~\bibnamefont {Nakatsuji}},\ and\ \bibinfo {author} {\bibfnamefont
  {Y.~C.}\ \bibnamefont {Otani}},\ }\bibfield  {title} {\bibinfo {title} {Giant
  field-like torque by the out-of-plane magnetic spin hall effect in a
  topological antiferromagnet},\ }\href
  {https://doi.org/10.1038/s41467-021-26453-y} {\bibfield  {journal} {\bibinfo
  {journal} {Nature Communications}\ }\textbf {\bibinfo {volume} {12}},\
  \bibinfo {pages} {1} (\bibinfo {year} {2021})}\BibitemShut {NoStop}%
\bibitem [{\citenamefont {Johnson}\ \emph {et~al.}(1996)\citenamefont
  {Johnson}, \citenamefont {Bloemen}, \citenamefont {den Broeder},\ and\
  \citenamefont {de~Vries}}]{Johnson1996Hani}%
  \BibitemOpen
  \bibfield  {author} {\bibinfo {author} {\bibfnamefont {M.~T.}\ \bibnamefont
  {Johnson}}, \bibinfo {author} {\bibfnamefont {P.~J.~H.}\ \bibnamefont
  {Bloemen}}, \bibinfo {author} {\bibfnamefont {F.~J.~A.}\ \bibnamefont {den
  Broeder}},\ and\ \bibinfo {author} {\bibfnamefont {J.~J.}\ \bibnamefont
  {de~Vries}},\ }\bibfield  {title} {\bibinfo {title} {Magnetic anisotropy in
  metallic multilayers},\ }\href {https://doi.org/10.1088/0034-4885/59/11/002}
  {\bibfield  {journal} {\bibinfo  {journal} {Reports on Progress in Physics}\
  }\textbf {\bibinfo {volume} {59}},\ \bibinfo {pages} {1409} (\bibinfo {year}
  {1996})}\BibitemShut {NoStop}%
\bibitem [{\citenamefont {Zayets}(2019)}]{ZayetsArch2019Hc}%
  \BibitemOpen
  \bibfield  {author} {\bibinfo {author} {\bibfnamefont {V.}~\bibnamefont
  {Zayets}},\ }\bibfield  {title} {\bibinfo {title} {Thermally activated
  magnetization reversal in a fecob nanomagnet. high-precision measurement
  method of coercive field, delta, retention time and size of nucleation
  domain},\ }\bibfield  {journal} {\bibinfo  {journal} {arXiv}\ }\textbf
  {\bibinfo {volume} {1908}},\ \href
  {https://doi.org/10.48550/ARXIV.1908.08435} {10.48550/ARXIV.1908.08435}
  (\bibinfo {year} {2019})\BibitemShut {NoStop}%
\bibitem [{\citenamefont {Slonczewski}(1996)}]{STT_Torque_Slonczewski_1996}%
  \BibitemOpen
  \bibfield  {author} {\bibinfo {author} {\bibfnamefont {J.}~\bibnamefont
  {Slonczewski}},\ }\bibfield  {title} {\bibinfo {title} {Current-driven
  excitation of magnetic multilayers},\ }\href
  {https://doi.org/10.1016/0304-8853(96)00062-5} {\bibfield  {journal}
  {\bibinfo  {journal} {Journal of Magnetism and Magnetic Materials}\ }\textbf
  {\bibinfo {volume} {159}},\ \bibinfo {pages} {L1} (\bibinfo {year}
  {1996})}\BibitemShut {NoStop}%
\bibitem [{\citenamefont {Berger}(1996)}]{STT_Torque_Berger_1996}%
  \BibitemOpen
  \bibfield  {author} {\bibinfo {author} {\bibfnamefont {L.}~\bibnamefont
  {Berger}},\ }\bibfield  {title} {\bibinfo {title} {Emission of spin waves by
  a magnetic multilayer traversed by a current},\ }\href
  {https://doi.org/10.1103/PhysRevB.54.9353} {\bibfield  {journal} {\bibinfo
  {journal} {Phys. Rev. B}\ }\textbf {\bibinfo {volume} {54}},\ \bibinfo
  {pages} {9353} (\bibinfo {year} {1996})}\BibitemShut {NoStop}%
\bibitem [{\citenamefont {Miron}\ \emph {et~al.}(2011)\citenamefont {Miron},
  \citenamefont {Garello}, \citenamefont {Gaudin}, \citenamefont {Zermatten},
  \citenamefont {Costache}, \citenamefont {Auffret}, \citenamefont {Bandiera},
  \citenamefont {Rodmacq}, \citenamefont {Schuhl},\ and\ \citenamefont
  {Gambardella}}]{SOT_Miron2011}%
  \BibitemOpen
  \bibfield  {author} {\bibinfo {author} {\bibfnamefont {I.~M.}\ \bibnamefont
  {Miron}}, \bibinfo {author} {\bibfnamefont {K.}~\bibnamefont {Garello}},
  \bibinfo {author} {\bibfnamefont {G.}~\bibnamefont {Gaudin}}, \bibinfo
  {author} {\bibfnamefont {P.~J.}\ \bibnamefont {Zermatten}}, \bibinfo {author}
  {\bibfnamefont {M.~V.}\ \bibnamefont {Costache}}, \bibinfo {author}
  {\bibfnamefont {S.}~\bibnamefont {Auffret}}, \bibinfo {author} {\bibfnamefont
  {S.}~\bibnamefont {Bandiera}}, \bibinfo {author} {\bibfnamefont
  {B.}~\bibnamefont {Rodmacq}}, \bibinfo {author} {\bibfnamefont
  {A.}~\bibnamefont {Schuhl}},\ and\ \bibinfo {author} {\bibfnamefont
  {P.}~\bibnamefont {Gambardella}},\ }\bibfield  {title} {\bibinfo {title}
  {Perpendicular switching of a single ferromagnetic layer induced by in-plane
  current injection},\ }\href {https://doi.org/10.1038/nature10309} {\bibfield
  {journal} {\bibinfo  {journal} {Nature}\ }\textbf {\bibinfo {volume} {476}},\
  \bibinfo {pages} {189} (\bibinfo {year} {2011})}\BibitemShut {NoStop}%
\bibitem [{\citenamefont {Zayets}(2018)}]{ZayetsJMMM2018Holes}%
  \BibitemOpen
  \bibfield  {author} {\bibinfo {author} {\bibfnamefont {V.}~\bibnamefont
  {Zayets}},\ }\bibfield  {title} {\bibinfo {title} {Spin transport of
  electrons and holes in a metal and in a semiconductor},\ }\href
  {https://doi.org/https://doi.org/10.1016/j.jmmm.2017.08.072} {\bibfield
  {journal} {\bibinfo  {journal} {Journal of Magnetism and Magnetic Materials}\
  }\textbf {\bibinfo {volume} {445}},\ \bibinfo {pages} {53} (\bibinfo {year}
  {2018})}\BibitemShut {NoStop}%
\bibitem [{\citenamefont {Shiota}\ \emph {et~al.}(2012)\citenamefont {Shiota},
  \citenamefont {Miwa}, \citenamefont {Nozaki}, \citenamefont {Bonell},
  \citenamefont {Mizuochi}, \citenamefont {Shinjo}, \citenamefont {Kubota},
  \citenamefont {Yuasa},\ and\ \citenamefont {Suzuki}}]{VCMA_Shiota2012}%
  \BibitemOpen
  \bibfield  {author} {\bibinfo {author} {\bibfnamefont {Y.}~\bibnamefont
  {Shiota}}, \bibinfo {author} {\bibfnamefont {S.}~\bibnamefont {Miwa}},
  \bibinfo {author} {\bibfnamefont {T.}~\bibnamefont {Nozaki}}, \bibinfo
  {author} {\bibfnamefont {F.}~\bibnamefont {Bonell}}, \bibinfo {author}
  {\bibfnamefont {N.}~\bibnamefont {Mizuochi}}, \bibinfo {author}
  {\bibfnamefont {T.}~\bibnamefont {Shinjo}}, \bibinfo {author} {\bibfnamefont
  {H.}~\bibnamefont {Kubota}}, \bibinfo {author} {\bibfnamefont
  {S.}~\bibnamefont {Yuasa}},\ and\ \bibinfo {author} {\bibfnamefont
  {Y.}~\bibnamefont {Suzuki}},\ }\bibfield  {title} {\bibinfo {title} {Pulse
  voltage-induced dynamic magnetization switching in magnetic tunneling
  junctions with high resistance-area product},\ }\href
  {https://doi.org/10.1063/1.4751035} {\bibfield  {journal} {\bibinfo
  {journal} {Applied Physics Letters}\ }\textbf {\bibinfo {volume} {101}},\
  \bibinfo {pages} {102406} (\bibinfo {year} {2012})}\BibitemShut {NoStop}%
\bibitem [{\citenamefont {Nozaki}\ \emph {et~al.}(2016)\citenamefont {Nozaki},
  \citenamefont {Kozio\l{}-Rachwa\l{}}, \citenamefont
  {Skowro\ifmmode~\acute{n}\else \'{n}\fi{}ski}, \citenamefont {Zayets},
  \citenamefont {Shiota}, \citenamefont {Tamaru}, \citenamefont {Kubota},
  \citenamefont {Fukushima}, \citenamefont {Yuasa},\ and\ \citenamefont
  {Suzuki}}]{VCMA_Nozaki2016}%
  \BibitemOpen
  \bibfield  {author} {\bibinfo {author} {\bibfnamefont {T.}~\bibnamefont
  {Nozaki}}, \bibinfo {author} {\bibfnamefont {A.}~\bibnamefont
  {Kozio\l{}-Rachwa\l{}}}, \bibinfo {author} {\bibfnamefont {W.}~\bibnamefont
  {Skowro\ifmmode~\acute{n}\else \'{n}\fi{}ski}}, \bibinfo {author}
  {\bibfnamefont {V.}~\bibnamefont {Zayets}}, \bibinfo {author} {\bibfnamefont
  {Y.}~\bibnamefont {Shiota}}, \bibinfo {author} {\bibfnamefont
  {S.}~\bibnamefont {Tamaru}}, \bibinfo {author} {\bibfnamefont
  {H.}~\bibnamefont {Kubota}}, \bibinfo {author} {\bibfnamefont
  {A.}~\bibnamefont {Fukushima}}, \bibinfo {author} {\bibfnamefont
  {S.}~\bibnamefont {Yuasa}},\ and\ \bibinfo {author} {\bibfnamefont
  {Y.}~\bibnamefont {Suzuki}},\ }\bibfield  {title} {\bibinfo {title} {Large
  voltage-induced changes in the perpendicular magnetic anisotropy of an
  mgo-based tunnel junction with an ultrathin fe layer},\ }\href
  {https://doi.org/10.1103/PhysRevApplied.5.044006} {\bibfield  {journal}
  {\bibinfo  {journal} {Phys. Rev. Applied}\ }\textbf {\bibinfo {volume} {5}},\
  \bibinfo {pages} {044006} (\bibinfo {year} {2016})}\BibitemShut {NoStop}%
\bibitem [{\citenamefont {Rippard}\ \emph
  {et~al.}(2004{\natexlab{a}})\citenamefont {Rippard}, \citenamefont {Pufall},
  \citenamefont {Kaka}, \citenamefont {Russek},\ and\ \citenamefont
  {Silva}}]{STO.PRL.1st}%
  \BibitemOpen
  \bibfield  {author} {\bibinfo {author} {\bibfnamefont {W.~H.}\ \bibnamefont
  {Rippard}}, \bibinfo {author} {\bibfnamefont {M.~R.}\ \bibnamefont {Pufall}},
  \bibinfo {author} {\bibfnamefont {S.}~\bibnamefont {Kaka}}, \bibinfo {author}
  {\bibfnamefont {S.~E.}\ \bibnamefont {Russek}},\ and\ \bibinfo {author}
  {\bibfnamefont {T.~J.}\ \bibnamefont {Silva}},\ }\bibfield  {title} {\bibinfo
  {title} {Direct-current induced dynamics in
  ${\mathrm{c}\mathrm{o}}_{90}{\mathrm{f}\mathrm{e}}_{10}/{\mathrm{n}\mathrm{i}}_{80}{\mathrm{f}\mathrm{e}}_{20}$
  point contacts},\ }\href {https://doi.org/10.1103/PhysRevLett.92.027201}
  {\bibfield  {journal} {\bibinfo  {journal} {Phys. Rev. Lett.}\ }\textbf
  {\bibinfo {volume} {92}},\ \bibinfo {pages} {027201} (\bibinfo {year}
  {2004}{\natexlab{a}})}\BibitemShut {NoStop}%
\bibitem [{\citenamefont {Rippard}\ \emph
  {et~al.}(2004{\natexlab{b}})\citenamefont {Rippard}, \citenamefont {Pufall},
  \citenamefont {Kaka}, \citenamefont {Silva},\ and\ \citenamefont
  {Russek}}]{STO.2004.PRB.2nd}%
  \BibitemOpen
  \bibfield  {author} {\bibinfo {author} {\bibfnamefont {W.~H.}\ \bibnamefont
  {Rippard}}, \bibinfo {author} {\bibfnamefont {M.~R.}\ \bibnamefont {Pufall}},
  \bibinfo {author} {\bibfnamefont {S.}~\bibnamefont {Kaka}}, \bibinfo {author}
  {\bibfnamefont {T.~J.}\ \bibnamefont {Silva}},\ and\ \bibinfo {author}
  {\bibfnamefont {S.~E.}\ \bibnamefont {Russek}},\ }\bibfield  {title}
  {\bibinfo {title} {Current-driven microwave dynamics in magnetic point
  contacts as a function of applied field angle},\ }\href
  {https://doi.org/10.1103/PhysRevB.70.100406} {\bibfield  {journal} {\bibinfo
  {journal} {Phys. Rev. B}\ }\textbf {\bibinfo {volume} {70}},\ \bibinfo
  {pages} {100406} (\bibinfo {year} {2004}{\natexlab{b}})}\BibitemShut
  {NoStop}%
\bibitem [{\citenamefont {Pufall}\ \emph {et~al.}(2005)\citenamefont {Pufall},
  \citenamefont {Rippard}, \citenamefont {Kaka}, \citenamefont {Silva},\ and\
  \citenamefont {Russek}}]{STO.NIST.2005}%
  \BibitemOpen
  \bibfield  {author} {\bibinfo {author} {\bibfnamefont {M.~R.}\ \bibnamefont
  {Pufall}}, \bibinfo {author} {\bibfnamefont {W.~H.}\ \bibnamefont {Rippard}},
  \bibinfo {author} {\bibfnamefont {S.}~\bibnamefont {Kaka}}, \bibinfo {author}
  {\bibfnamefont {T.~J.}\ \bibnamefont {Silva}},\ and\ \bibinfo {author}
  {\bibfnamefont {S.~E.}\ \bibnamefont {Russek}},\ }\bibfield  {title}
  {\bibinfo {title} {{Frequency modulation of spin-transfer oscillators}},\
  }\href {https://doi.org/10.1063/1.1875762} {\bibfield  {journal} {\bibinfo
  {journal} {Applied Physics Letters}\ }\textbf {\bibinfo {volume} {86}},\
  \bibinfo {pages} {082506} (\bibinfo {year} {2005})}\BibitemShut {NoStop}%
\bibitem [{\citenamefont {Dussaux}\ \emph {et~al.}(2017)\citenamefont
  {Dussaux}, \citenamefont {Georges}, \citenamefont {Grollier}, \citenamefont
  {Cros}, \citenamefont {Khvalkovskiy}, \citenamefont {Fukushima},
  \citenamefont {Konoto}, \citenamefont {Kubota}, \citenamefont {Yakushiji},
  \citenamefont {Yuasa}, \citenamefont {Zvezdin}, \citenamefont {Ando},\ and\
  \citenamefont {Fert}}]{STO.AIST.FERT.2010.basic}%
  \BibitemOpen
  \bibfield  {author} {\bibinfo {author} {\bibfnamefont {A.}~\bibnamefont
  {Dussaux}}, \bibinfo {author} {\bibfnamefont {B.}~\bibnamefont {Georges}},
  \bibinfo {author} {\bibfnamefont {J.}~\bibnamefont {Grollier}}, \bibinfo
  {author} {\bibfnamefont {V.}~\bibnamefont {Cros}}, \bibinfo {author}
  {\bibfnamefont {A.}~\bibnamefont {Khvalkovskiy}}, \bibinfo {author}
  {\bibfnamefont {A.}~\bibnamefont {Fukushima}}, \bibinfo {author}
  {\bibfnamefont {M.}~\bibnamefont {Konoto}}, \bibinfo {author} {\bibfnamefont
  {H.}~\bibnamefont {Kubota}}, \bibinfo {author} {\bibfnamefont
  {K.}~\bibnamefont {Yakushiji}}, \bibinfo {author} {\bibfnamefont
  {S.}~\bibnamefont {Yuasa}}, \bibinfo {author} {\bibfnamefont
  {K.}~\bibnamefont {Zvezdin}}, \bibinfo {author} {\bibfnamefont
  {K.}~\bibnamefont {Ando}},\ and\ \bibinfo {author} {\bibfnamefont
  {A.}~\bibnamefont {Fert}},\ }\bibfield  {title} {\bibinfo {title} {Large
  microwave generation from current-driven magnetic vortex oscillators in
  magnetic tunnel junctions},\ }\href {https://doi.org/10.1038/ncomms1006}
  {\bibfield  {journal} {\bibinfo  {journal} {Nature Communications}\ }\textbf
  {\bibinfo {volume} {1}},\ \bibinfo {pages} {8} (\bibinfo {year}
  {2017})}\BibitemShut {NoStop}%
\bibitem [{\citenamefont {Kumar}\ \emph {et~al.}(2016)\citenamefont {Kumar},
  \citenamefont {Konishi}, \citenamefont {Kumar}, \citenamefont {Miwa},
  \citenamefont {Fukushima}, \citenamefont {Yakushiji}, \citenamefont {Yuasa},
  \citenamefont {Kubota}, \citenamefont {Tomy}, \citenamefont {Prabhakar},
  \citenamefont {Suzuki},\ and\ \citenamefont
  {Tulapurkar}}]{STO.AIST.2016.feedback.magnetic.field}%
  \BibitemOpen
  \bibfield  {author} {\bibinfo {author} {\bibfnamefont {D.}~\bibnamefont
  {Kumar}}, \bibinfo {author} {\bibfnamefont {K.}~\bibnamefont {Konishi}},
  \bibinfo {author} {\bibfnamefont {N.}~\bibnamefont {Kumar}}, \bibinfo
  {author} {\bibfnamefont {S.}~\bibnamefont {Miwa}}, \bibinfo {author}
  {\bibfnamefont {A.}~\bibnamefont {Fukushima}}, \bibinfo {author}
  {\bibfnamefont {K.}~\bibnamefont {Yakushiji}}, \bibinfo {author}
  {\bibfnamefont {S.}~\bibnamefont {Yuasa}}, \bibinfo {author} {\bibfnamefont
  {H.}~\bibnamefont {Kubota}}, \bibinfo {author} {\bibfnamefont {C.~V.}\
  \bibnamefont {Tomy}}, \bibinfo {author} {\bibfnamefont {A.}~\bibnamefont
  {Prabhakar}}, \bibinfo {author} {\bibfnamefont {Y.}~\bibnamefont {Suzuki}},\
  and\ \bibinfo {author} {\bibfnamefont {A.}~\bibnamefont {Tulapurkar}},\
  }\bibfield  {title} {\bibinfo {title} {Coherent microwave generation by
  spintronic feedback oscillator},\ }\href {https://doi.org/10.1038/srep30747}
  {\bibfield  {journal} {\bibinfo  {journal} {Scientific Reports}\ }\textbf
  {\bibinfo {volume} {6}},\ \bibinfo {pages} {30747} (\bibinfo {year}
  {2016})}\BibitemShut {NoStop}%
\bibitem [{\citenamefont {Tsunegi}\ \emph {et~al.}(2016)\citenamefont
  {Tsunegi}, \citenamefont {Grimaldi}, \citenamefont {Lebrun}, \citenamefont
  {Kubota}, \citenamefont {Jenkins}, \citenamefont {Yakushiji}, \citenamefont
  {Fukushima}, \citenamefont {Bortolotti}, \citenamefont {Grollier},
  \citenamefont {Yuasa},\ and\ \citenamefont
  {Cros}}]{STO.AIST.2016.delayFeedback}%
  \BibitemOpen
  \bibfield  {author} {\bibinfo {author} {\bibfnamefont {S.}~\bibnamefont
  {Tsunegi}}, \bibinfo {author} {\bibfnamefont {E.}~\bibnamefont {Grimaldi}},
  \bibinfo {author} {\bibfnamefont {R.}~\bibnamefont {Lebrun}}, \bibinfo
  {author} {\bibfnamefont {H.}~\bibnamefont {Kubota}}, \bibinfo {author}
  {\bibfnamefont {A.~S.}\ \bibnamefont {Jenkins}}, \bibinfo {author}
  {\bibfnamefont {K.}~\bibnamefont {Yakushiji}}, \bibinfo {author}
  {\bibfnamefont {A.}~\bibnamefont {Fukushima}}, \bibinfo {author}
  {\bibfnamefont {P.}~\bibnamefont {Bortolotti}}, \bibinfo {author}
  {\bibfnamefont {J.}~\bibnamefont {Grollier}}, \bibinfo {author}
  {\bibfnamefont {S.}~\bibnamefont {Yuasa}},\ and\ \bibinfo {author}
  {\bibfnamefont {V.}~\bibnamefont {Cros}},\ }\bibfield  {title} {\bibinfo
  {title} {Self-injection locking of a vortex spin torque oscillator by delayed
  feedback},\ }\href {https://doi.org/10.1038/srep26849} {\bibfield  {journal}
  {\bibinfo  {journal} {Scientific Reports}\ }\textbf {\bibinfo {volume} {6}},\
  \bibinfo {pages} {26849} (\bibinfo {year} {2016})}\BibitemShut {NoStop}%
\bibitem [{\citenamefont {Singh}\ \emph {et~al.}(2018)\citenamefont {Singh},
  \citenamefont {Bose}, \citenamefont {Bhuktare}, \citenamefont {Fukushima},
  \citenamefont {Yakushiji}, \citenamefont {Yuasa}, \citenamefont {Kubota},\
  and\ \citenamefont {Tulapurkar}}]{STO.AIST.2018.feedback.magnetic.field}%
  \BibitemOpen
  \bibfield  {author} {\bibinfo {author} {\bibfnamefont {H.}~\bibnamefont
  {Singh}}, \bibinfo {author} {\bibfnamefont {A.}~\bibnamefont {Bose}},
  \bibinfo {author} {\bibfnamefont {S.}~\bibnamefont {Bhuktare}}, \bibinfo
  {author} {\bibfnamefont {A.}~\bibnamefont {Fukushima}}, \bibinfo {author}
  {\bibfnamefont {K.}~\bibnamefont {Yakushiji}}, \bibinfo {author}
  {\bibfnamefont {S.}~\bibnamefont {Yuasa}}, \bibinfo {author} {\bibfnamefont
  {H.}~\bibnamefont {Kubota}},\ and\ \bibinfo {author} {\bibfnamefont {A.~A.}\
  \bibnamefont {Tulapurkar}},\ }\bibfield  {title} {\bibinfo {title}
  {Self-injection locking of a spin torque nano-oscillator to magnetic field
  feedback},\ }\href {https://doi.org/10.1103/PhysRevApplied.10.024001}
  {\bibfield  {journal} {\bibinfo  {journal} {Phys. Rev. Appl.}\ }\textbf
  {\bibinfo {volume} {10}},\ \bibinfo {pages} {024001} (\bibinfo {year}
  {2018})}\BibitemShut {NoStop}%
\end{thebibliography}%

\appendix
\label{AppendixLL}

\section{Calculation of the parametric torque from Landau-Lifshitz equation }

In the subsequent analysis, the Landau-Lifshitz (LL) equation is solved under the influence of parametric pumping generated by an external magnetic field oscillating at a frequency near the Ferromagnetic Resonance (FMR) frequency. The objective is to determine the pumping torque resulting from this parametric pumping. The LL equation, including the precession and parametric terms but excluding damping, can be expressed as follows:

\begin{equation}
{{\partial \vec m} \over {\partial t}} =  - \gamma \vec m \times \left( {\vec H + \vec H^{(CI)} } \right)
\end{equation}

where $\gamma$ is the electron gyromagnetic ratio,  $H$ is applied along the z- axis, $H^{(CI)}$ is applied along the x- axis and is oscillating with the frequency $\omega$:

\begin{equation}
H^{(CI)}  = H_x^{(CI)} \sin \left( {\omega t} \right)
\label{Hcisin}
\end{equation}

The scalar form of the Eq. \ref{Hcisin} is

\begin{equation}
\begin{array}{l}
	{{\partial m_x } \over {\partial t}} =  - \omega _L m_y 
	\\
	{{{\partial m_y } \over {\partial t}} = \omega _L m_x  - \omega _C   m_z   \sin \left( {\omega t} \right)}
	\\
	{{\partial m_z } \over {\partial t}} =  - \omega _C m_y \sin \left( {\omega t} \right)
\end{array}
\label{LL1}
\end{equation}

where the Larmor frequencies $\omega _L  = \gamma H_z$ and $\omega _C  = \gamma H_x^{(CI)}$.

Introduction of new unknowns 
\begin{equation}
\begin{array}{ll}
	{m_ +   = m_x  + i \cdot m_y } & {m_ -   = m_x  - i \cdot m_y }			
\end{array}
\label{m+}
\end{equation}

and addition/ subtraction of the 1st and 2nd equations of \ref{LL1} give

\begin{equation}
\begin{array}{l}
	{{{\partial m_ +  } \over {\partial t}} = i\omega _L m_ +   - i \cdot \omega _C  \cdot m_z  \cdot \sin \left( {\omega t} \right)}\\
	
	{{{\partial m_ -  } \over {\partial t}} =  - i\omega _L m_ -   + i \cdot \omega _C  \cdot m_z  \cdot \sin \left( {\omega t} \right)}	\\
	
	{{{\partial m_z } \over {\partial t}} = \omega _C {{m_ +   - m_ -  } \over {2i}} \cdot \sin \left( {\omega t} \right)}	
	\label{LL2}	
\end{array}
\end{equation}

The solution of Eq. \ref{LL2} can be found in form:
\begin{equation}
\begin{array}{l}
	{m_ +  \left( t \right) = M \cdot \sin \left( {\theta \left( t \right)} \right) \cdot e^{i\phi _ +  \left( t \right)} }  \\ 
	{m_ -  \left( t \right) = M \cdot \sin \left( {\theta \left( t \right)} \right) \cdot e^{ - i\phi _ -  \left( t \right)} }  \\
	{m_z \left( t \right) = M \cdot \cos \left( {\theta \left( t \right)} \right)}
	\label{LL_sol1}
\end{array}
\end{equation}

where $\theta$, $\phi_+$ and $\phi_-$  are new unknowns and $M$ is the magnetization.

Substitution of \ref{LL_sol1} into \ref{LL2} gives

\begin{equation}
\begin{array}{l}
	{{{\partial \theta } \over {\partial t}} + i \cdot \tan \left( \theta  \right)\left[ {{{\partial \phi _ +  } \over {\partial t}} - \omega _L } \right] =  - i \cdot \omega _C  \cdot \sin \left( {\omega t} \right) \cdot e^{ - i\phi _ +  } }  \\
	{{{\partial \theta } \over {\partial t}} - i \cdot \tan \left( \theta  \right)\left[ {{{\partial \phi _ -  } \over {\partial t}} - \omega _L } \right] =  + i \cdot \omega _C  \cdot \sin \left( {\omega t} \right)e^{i\phi _ -  } }  \\
	{{{\partial \theta } \over {\partial t}} =  - \omega _C {{e^{i\phi _ +  }  - e^{ - i\phi _ -  } } \over {2i}} \cdot \sin \left( {\omega t} \right)}
\end{array}
\label{LL3}
\end{equation}

Substitution of the 3rd Eq. into the 1st and 2nd Eqs. of \ref{LL3} gives

\begin{equation}
\begin{array}{l}
	\tan \left( \theta  \right)\left[ {{{\partial \phi _ +  } \over {\partial t}} - \omega _L } \right] = \omega _C  \cdot \left( { - e^{ - i\phi _ +  }  - {{e^{i\phi _ +  }  - e^{ - i\phi _ -  } } \over 2}} \right) \cdot \sin \left( {\omega t} \right)  \\
	
	\tan \left( \theta  \right)\left[ {{{\partial \phi _ -  } \over {\partial t}} - \omega _L } \right] = \omega _C  \cdot \left( { - e^{i\phi _ -  }  + {{e^{i\phi _ +  }  - e^{ - i\phi _ -  } } \over 2}} \right) \cdot \sin \left( {\omega t} \right) \\ 
	
	{{\partial \theta } \over {\partial t}} =  - \omega _C {{e^{i\phi _ +  }  - e^{ - i\phi _ -  } } \over {2i}} \cdot \sin \left( {\omega t} \right)  
	\label{LL3}
\end{array}
\end{equation}

Taking into account that the oscillating magnetic field is small

\begin{equation}
\begin{array}{ll}
	{H_{x,\omega }  \ll H} & {\omega _C   \ll \omega _L }

\end{array}
\end{equation}

The solution of Eqs. \ref{LL3} can be expanded into a Taylor series, where $\omega_C$ is as a small parameter:

\begin{equation}
\begin{array}{l}
	\phi _ +   = \phi _{ + ,0}  + \omega _C  \cdot \phi _{ + ,1}   \\ 
	\phi _ -   = \phi _{ - ,0}  + \omega _C  \cdot \phi _{ - ,1}   \\ 
	\theta  = \theta _0  + \omega _C  \cdot \theta _1  
\end{array}
\label{Taylor}
\end{equation}

Substitution of \ref{LL3} into \ref{Taylor} gives

\begin{equation}
\begin{array}{l}
	\left[ {\tan \left( {\theta _0 } \right) + \omega _C  \cdot \theta _1  + \omega _C  \cdot \theta _1  \cdot \tan \left( {\theta _0 } \right)^2 } \right] \cdot \\
	\cdot  \left[ {{{\partial \phi _{ + ,0} } \over {\partial t}} - \omega _L  + \omega _C {{\partial \phi _{ + ,1} } \over {\partial t}}} \right] = \\
	=	 \omega_C  \cdot \left( { - e^{ - i\phi _{ + ,0} }  - {{e^{i\phi _{ + ,0} }  - e^{ - i\phi _{ - ,0} } } \over 2}} \right) \cdot \sin \left( {\omega t} \right)  \\ 	
	\left[ {\tan \left( {\theta _0 } \right) + \omega _C  \cdot \theta _1  + \omega _C  \cdot \theta _1  \cdot \tan \left( {\theta _0 } \right)^2 } \right] \cdot \\
	\cdot  \left[ {{{\partial \phi _{ - ,0} } \over {\partial t}} - \omega _L  + \omega _C {{\partial \phi _{ - ,1} } \over {\partial t}}} \right] = \\
	= \omega _C  \cdot \left( { - e^{i\phi _{ - ,0} }  + {{e^{i\phi _{ + ,0} }  - e^{ - i\phi _{ - ,0} } } \over 2}} \right) \cdot \sin \left( {\omega t} \right)  \\ 
	
	{{{\partial \theta _0 } \over {\partial t}} + \omega _C {{\partial \theta _1 } \over {\partial t}} =  - \omega _C {{e^{i\phi _{ + ,0} }  - e^{ - i\phi _{ - ,0} } } \over {2i}} \cdot \sin \left( {\omega t} \right)} 
\end{array}
\label{LL4}
\end{equation}

Comparison of parts, which are proportional to $(\omega_C)^0$, gives

\begin{equation}
\begin{array}{l}
	{\tan \left( {\theta _0 } \right)\left[ {{{\partial \phi _{0, + } } \over {\partial t}} - \omega _L } \right] = 0}  \\
	{\tan \left( {\theta _0 } \right)\left[ {{{\partial \phi _{0, - } } \over {\partial t}} - \omega _L } \right] = 0}  \\ 
	{{{\partial \theta _0 } \over {\partial t}} = 0}  
	\label{Solution0}
\end{array}
\end{equation}

The solution of Eqs.\ref{Solution0} gives a static precession at the Larmor frequency $\omega_L$ and at constant precession angle $\theta_0$ as

\begin{equation}
\begin{array}{l}
	\phi _{ + ,0}  = \omega _L t + \varphi   \\ 
	\phi _{ - ,0}  = \omega _L t + \varphi   \\ 
	\theta _0  = const  
\end{array}
\label{Solution0a}
\end{equation}

where $\varphi$ is the phase of the precession oscillation with respect to the phase of the parametric pumping. This zero- approximation describes the case of the magnetization precession without the parametric pumping. Substitution of \ref{Solution0a} into \ref{m+} and  \ref{Solution0} gives the temporal evolution for the precession at constant precession angle as

\begin{equation}
\begin{array}{l}
	m_x \left( t \right) = M \cdot \sin \left( {\theta _0 } \right) \cdot \cos \left( {\omega _L t + \varphi } \right)  \\
	
	m_ -  \left( t \right) = M \cdot \sin \left( {\theta _0 } \right) \cdot \sin \left( {\omega _L t + \varphi } \right)  \\
	
	m_z \left( t \right) = M \cdot \cos \left( {\theta _0 } \right)
\end{array}
\end{equation}

Comparison of parts in \ref{LL4}, which are proportional to  $(\omega_C)^1$, gives

\begin{equation}
\begin{array}{l}
	\tan \left( {\theta _0 } \right){{\partial \phi _{ + ,1} } \over {\partial t}} = \left( { - e^{ - i\left( {\omega _L t + \varphi } \right)}  - \cos \left( {\omega _L t + \varphi } \right)} \right) \cdot \sin \left( {\omega t} \right)  \\ 
	\tan \left( {\theta _0 } \right){{\partial \phi _{ + ,1} } \over {\partial t}} = \left( { - e^{i\left( {\omega _L t + \varphi } \right)}  + \cos \left( {\omega _L t + \varphi } \right)} \right) \cdot \sin \left( {\omega t} \right)  \\ 
	{{\partial \theta _1 } \over {\partial t}} =  - \sin \left( {\omega _L t + \varphi } \right) \cdot \sin \left( {\omega t} \right)  
\end{array}
\label{Solution1}
\end{equation}

The solution of the third equation of Eqs.\ref{Solution1} is

\begin{equation}
2 \theta_1=\frac{\sin([\omega _L  - \omega]t+\varphi)}{\omega _L  - \omega}+
\frac{\sin([\omega _L  + \omega]t+\varphi)}{\omega _L  + \omega}
\label{Solution1a}
\end{equation}

Ignoring the fast oscillating part at the frequency $\omega+\omega_L$, substitution of \ref{Solution1a} into \ref{Taylor} gives the temporal evolution of the precession angle as

\begin{equation}
\theta  = \theta _0  - {{\omega _C } \over 2} \cdot \frac{\sin([\omega _L  - \omega]t+\varphi)}{\omega _L  - \omega}
\label{SolutionAngle}
\end{equation}

Differentiation of Eq. \ref{SolutionAngle}  gives the parametric torque as

\begin{equation}
\frac{\partial \theta}{\partial t}=
-{{\omega _C } \over 2} \cdot \cos ([\omega_L-\omega]t+\varphi)
\label{torque}
\end{equation}

In case when the frequency and phase of the parametric pumping is fully synchronized with the precession frequency and phase:

\begin{equation}
\begin{array}{ll}
	{\omega=\omega_L} & {\varphi=\pi}
\end{array}
\end{equation}

the parametric torque is calculated from Eq. \ref{torque} as

\begin{equation}
\frac{\partial \theta}{\partial t}=
{{\omega _C } \over 2}={{\gamma } \over 2}  H_x^{(CI)}
\end{equation}

Assuming that the MR ratio $k_{MR}$ is small:
\begin{equation}
k_{MR} \sin(\theta)<<1
\end{equation}

Eq. \ref{H_CI} can be simplified as
\begin{equation}
\begin{array}{l}
	{H^{(CI)}} = 
	\frac{{\varsigma  \cdot {I_{DC}}}}{{1 - {k_{MR}}
			\sin \left( \theta  \right) \cdot \sin \left( {\omega t} \right)}}
	\approx \\
	\approx	\varsigma  \cdot I_{DC} (1+ k_{MR} \sin(\theta) \sin(\omega t) )
\end{array}	
\label{H_CI2}
\end{equation}

$H_x^{(CI)} $ is the oscillating part of $H^{(CI)}$ (See Eq.\ref{Hcisin}). Substitution of Eq. \ref{H_CI2} gives the parametric torque as

\begin{equation}
{\left( {\frac{{\partial \theta }}{{\partial t}}} \right)_{param}}=
{\gamma \cdot	\varsigma \cdot  I_{DC} \cdot  k_{MR} \over 2 } \cdot \sin(\theta) 
\end{equation}

It is worth noting that the parametric torque changes its polarity when the direction of the DC current is reversed, leading to magnetization reversal in the opposite direction.

\begin{figure}[t]
\begin{center}
	\includegraphics[width=7 cm]{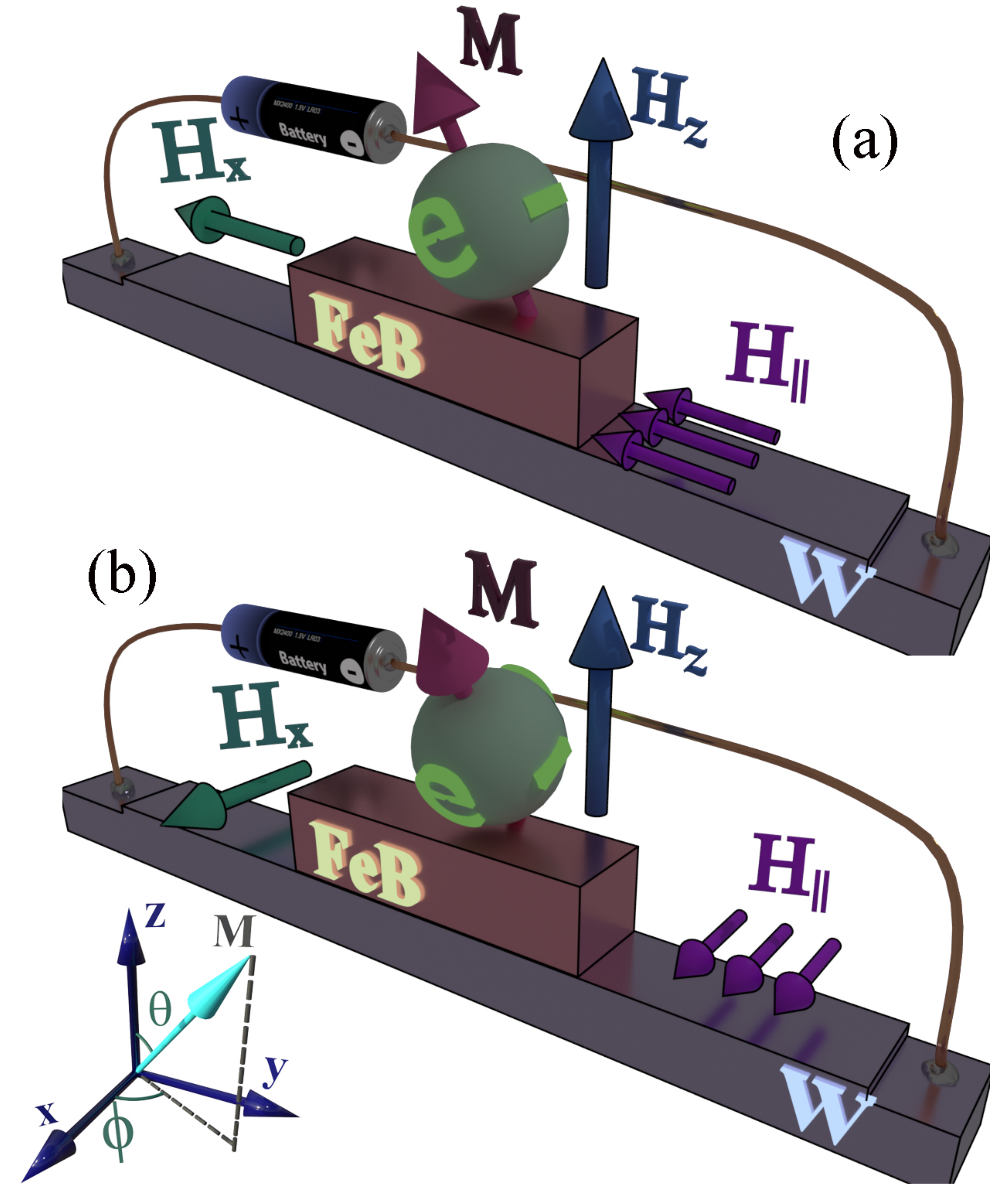}
\end{center}
\caption{\label{fig:figExp1} 
	Experimental setup for measuring magnetic fields generated by spin accumulation. An external magnetic field  $H_z$  is applied along the easy axis of the FeB nanomagnet and used as a parameter. An additional magnetic field $H_x$ is scanned perpendicular to the easy axis, in two configurations: (a) parallel to the current and (b) perpendicular to the current direction. $H_||$ represents a measured component of the magnetic field produced by the spin accumulation. $M$ denotes the magnetization. 
}
\end{figure}


\section{ Details of fabrication and measurement }
\label{AppendixExpSetup}

The sample was prepared on a $Si$/$SiO_2$ substrate. A $W$(3 nm)/$FeB$(1.2 nm)/$MgO$(7 nm)/$Ta$(1 nm)/$Ru$ (5 nm) film was sputtering at room temperature and subsequently  annealed at $T=250~0 C$. The $FeB$ was covered by a $MgO$ (10 nm) layer in order to induce a sufficient Perpendicular Magnetic Anisotropy (PMA ). A $Ta$/$Ru$ layer was used to protect the $MgO$ from exposure to air. The fabrication process involved several nanofabrication steps. Each step utilized either electron-beam (EB) or optical lithography, followed by $Ar$ milling and lift-off procedures. The alignment for each subsequent fabrication step was maintained within 5 nm precision.  Etching material was monitored in-situ by a Secondary-ion-mass-spectroscopy (SIMS) detector to ensure the required precision of etching depth. 

\begin{figure}[t]
	\begin{center}
		\includegraphics[width=5.5 cm]{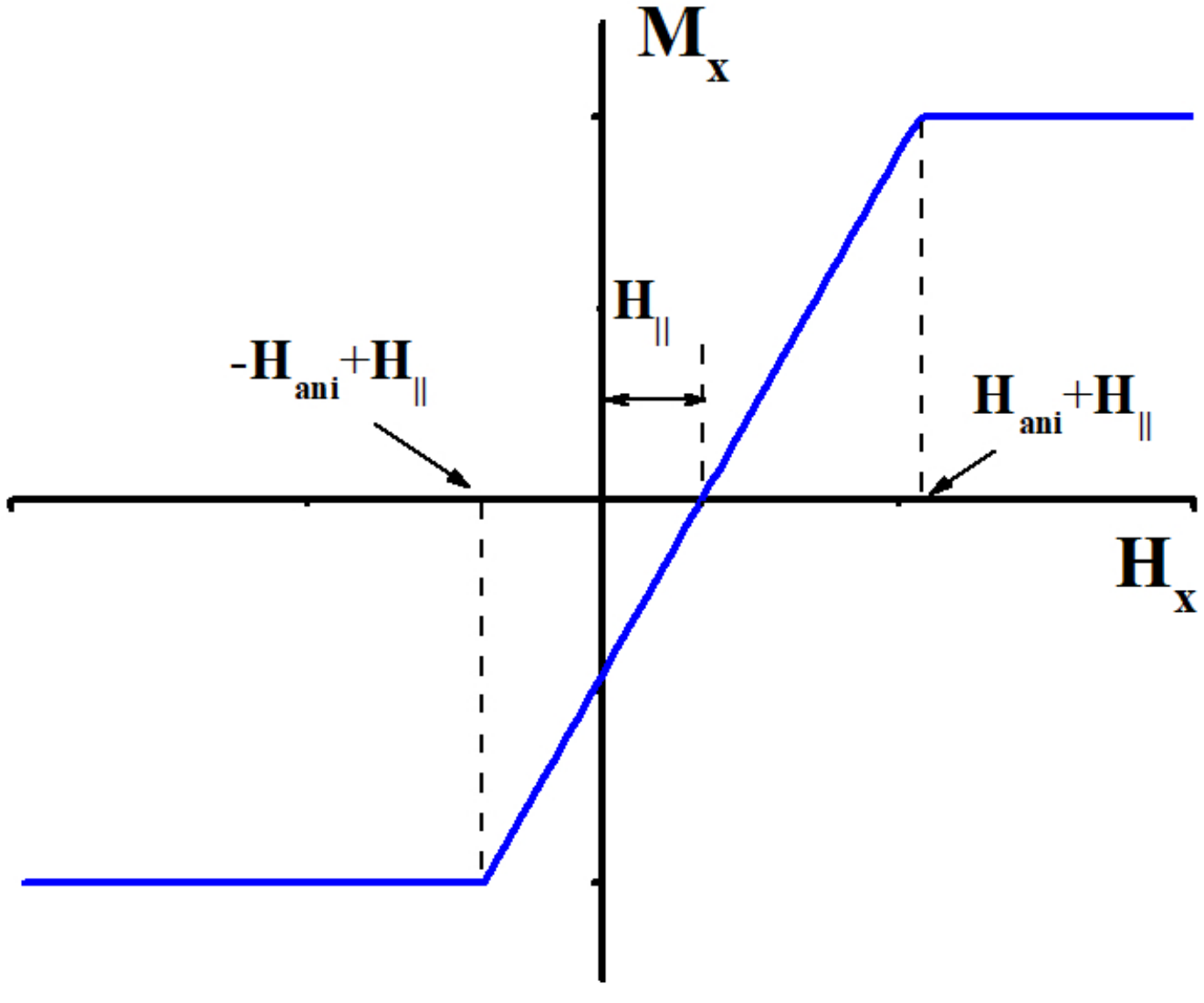}
	\end{center}
	\caption{\label{fig:figExp2} 
		Measurement principle. Schematic diagram. The in-plane component of magnetization $M_x$ vs. scanned in-plane magnetic field $H_x$. The dependence is asymmetric versus a reversal of $H_x$ due to existence of  the intrinsic magnetic field $H_{||}$. $H_{||}$ includes the current- induced magnetic field $H^{(CI)}$. The $H_{||}$ is evaluated by minimizing difference between positive and negative parts of the dependence with an offset. 
	}
\end{figure}

Nanomagnets of various sizes, ranging from 50 nm x 50 nm to 2000 nm x 2000 nm, were fabricated at different places of one wafer. The etching of nanomagnets stopped at the top of the $W$ layer. Subsequently, the $W$ nanowire with a $SiO_2$ (100 nm) isolation layer was fabricated. A pair of Hall probes was precisely aligned with the position of the nanomagnet, and the width of the nanowire matched the width of the corresponding nanomagnet. The etching process of the nanowire halted at the top of the $SiO_2$ substrate. Finally, we fabricated the $Cr$(2 nm)/$Au$(200 nm) contacts in the etched $SiO_2$, with the etching process of the contact ending in the middle of the $W$ layer.

The measurements were conducted at room temperature, significantly below the Curie temperature of $FeB$. The Hall voltage was measured by a nanovoltmeter, while the electrical current was both supplied and measured with a current source. The magnetic field was generated by an electromagnet, providing the capability to align the magnetic field in any direction. Consequently, the in-plane and perpendicular- to- plane components of the magnetic field  are controlled individually. The electromagnet was calibrated using Hall measurements performed on non-magnetic $Ta$, $W$, and $Ru$ nanowires. An intrinsic magnetic field was not detected in any of these non-magnetic nanowires.

The measurement process involved recording the Hall angle $\alpha_{Hall}$ during the scanning of an in-plane external magnetic field $H_x$ in two opposite directions. The perpendicular-to-plane magnetic field was employed as a parameter. Two consecutive measurements, with $H_x$ directed either along (fig. \ref*{fig:figExp1}(a))  or perpendicular to the current (fig. \ref*{fig:figExp1}(b)), were conducted.

Figure \ref*{fig:figExp2} demonstrates that the in-plane magnetization component, $M_x$, is linearly proportional to $H_x$. The field, at which the nanomagnet's magnetization aligns in-plane is referred to as the anisotropy field $H_{ani}$. It's crucial to note that the $M_x$ vs. $H_x$ relationship is not symmetrical due to the presence of the intrinsic in-plane magnetic field $H_{||}$. Consequently, the observed relationship between the Hall angle $\alpha_{Hall}$ and $H_x$ is also asymmetrical. To precisely evaluate $H_{||}$, the positive and negative segments of the measured relationship are jointly fitted. $H_{||}$ was determined by minimizing the following integral:

\begin{equation}
\int
(\alpha_{Hall}(H_x-H_{||})-\alpha_{Hall}(-[H_x-H_{||}]))^2 dH_x
\end{equation}

where $H_{||}$ is used as a minimization parameter.

In an ideal case, when there is no measurement noise, the integral equals zero at its minimum.

\end{document}